\documentclass[aas,appendixfloats]{emulateapj} 
\usepackage{natbib} 
\usepackage{url} 
\usepackage[usenames,dvipsnames]{color} 
\usepackage{graphicx} 
\usepackage{multirow}
\usepackage{enumitem}
\usepackage{textcomp}
\usepackage{wasysym}
\setlist{noitemsep} 
\slugcomment{version 0.5} 
\shorttitle{Unusually bright single pulses from the binary pulsar Ter5A} 
\shortauthors{A.~Bilous et al.}

\newcommand{\Msol}{M_{\astrosun}}

\newcommand{\T}{Ter5A}
\newcommand{\phiB}{\phi_\mathrm{B}}

\begin{document} 
\title{Unusually bright single pulses from the binary pulsar B1744$-$24A: a case of strong lensing?}

\author{A.~V.~Bilous\altaffilmark{1,2}}
\author{S.~M.~Ransom\altaffilmark{3}}
\author{P.~Demorest\altaffilmark{4}}

\altaffiltext{1}{Anton Pannekoek Institute for Astronomy, 
                 University of Amsterdam, Science Park 904, 
                 1098 XH Amsterdam, The Netherlands, \email{A.Bilous@uva.nl}}
\altaffiltext{2}{Department of Astronomy, University of Virginia, PO
  Box 400325, Charlottesville, VA 22904, USA} 
\altaffiltext{3}{National
  Radio Astronomy Observatory, Charlottesville, VA 22903, USA}
\altaffiltext{4}{National Radio Astronomy Observatory, PO Box O,
Socorro, NM 87801, USA  }
 
\begin{abstract} 
We present a study of unusually bright single pulses (BSPs) from a millisecond pulsar in an ablating binary system, 
B1744$-$24A, based on several multi-orbit observations with the Green Bank Telescope. These pulses come predominantly 
in time near eclipse ingress and egress, have intensities up to 40 times the average pulse intensity, and pulse widths 
similar to that of the average pulse profile. The average intensity, spectral index of radio emission, and the dispersion 
measure do not vary in connection with BSP outbursts. The average profile obtained  from BSPs has the same shape as the 
average profile from all pulses.  These properties make it difficult to explain BSPs via scintillation in the interstellar 
medium, as a separate emission mode, or as conventional giant pulses. BSPs from B1744$-$24A have similar properties to the strong 
pulses observed from the Black Widow binary pulsar B1957$+$10, which were recently attributed to strong lensing by the 
intrabinary material (Main et al.~2018). We argue that the strong lensing likely occurs in B1744$-$24A as well. 
For this system, the sizes and locations of the lenses are not well constrained by simple 1D
lensing models from Cordes et al.~(2017) and Main et al.~(2018).
This partly stems from the poor knowledge of several important physical parameters of the system.
\end{abstract}

\keywords{} 
 
\maketitle
\bigskip

\section{Introduction}

PSR~B1744$-$24A (hereafter \T) was the second eclipsing millisecond pulsar ever discovered \citep{Lyne1990}. Located in globular cluster 
Terzan 5, this 11.56\,ms pulsar is in a compact binary system with a relatively lightweight ($0.089\Msol$) companion, presumably a main 
sequence star or a white dwarf \citep{Nice1992}.

\T's radio eclipses are highly variable, lasting from approximately a quarter of an orbit to several full orbits \citep{Nice1992}. Eclipses 
are usually centered around inferior conjunction (at orbital phase or mean anomaly of 0.25), although short-duration ``mini-eclipses'' 
are sometimes detected at other orbital phases, and occasionally pulsar emission (albeit dispersed and attenuated) is visible throughout 
the inferior conjunction \citep{Thorsett1991}. This erratic behavior led to the suggestion that \T\ is ablating material from its companion 
and producing a spatially complex and dynamic stellar wind. The hydrodynamics of the outflow was explored by \citet{Tavani1991} and 
\citet{Rasio1991} and several plausible eclipse mechanisms have been proposed \citep[e.g.][]{Rasio1991,Thompson1994,Luo1995}.

\begin{table*}
\begin{center} 
\caption{Observing summary. The columns are: observing date, MJD at the start of the session, session duration, name of the band, central 
frequency, bandwidth, number of channels, sampling time, DM measured away from eclipses, and the average uneclipsed flux density.\label{table:obssum}}
\begin{tabular}{ccccccccccc} 
\hline\\ 
\parbox{1.2cm}{\centering  Session \textit{yymmdd}}&
\parbox{1.2cm}{\centering  Start MJD}&
\parbox{1.2cm}{\centering Duration (hr)} &
\parbox{1.2cm}{\centering  Band name}&
\parbox{1.2cm}{\centering $\nu_c$ (MHz)} &
\parbox{1.2cm}{\centering BW (MHz)} & 
\parbox{1.2cm}{\centering \# of channels} &
\parbox{1.2cm}{\centering Sampling time \\($\mu$s)} & 
\parbox{1.8cm}{\centering DM \\(pc cm$^{-3}$)} &
\parbox{1.9cm}{\centering Average uneclipsed flux density (mJy)} 
\\ [0.3cm]
\hline\\
100815 & 55422.9 & 7.7 & S & 1999.2 & 800 & 512 & 10.24 & 242.300(2)\hphantom{0} &  1.2  \\
101002 & 55472.8 & 7.5 & UHF & \hphantom{0}820.2 & 200 & 512 & 10.24 & 242.3332(3) & 8.1 \\
110925 & 55829.8 & 7.7 & L & 1499.2 & 800 & 512 & 10.24 & 242.3362(3) & 2.2 \\
120415 & 56032.2 & 4.2 & S & 1999.2 & 800 & 512 & 10.24 & 242.339(4)\hphantom{0} & 2.4 \\
121006 & 56206.8 & 6.7 & L & 1499.2 & 800 & 512 & 10.24 & 242.358(1)\hphantom{0} & 2.4 \\
131022 & 56587.7 & 7.7 & L & 1499.2 & 800 & 512 & 10.24 & 242.3617(2) & 4.9 \\
140114 & 56671.5 & 7.6 & L & 1499.2 & 800 & 512 & 10.24 & 242.3609(4) & 5.8 \\
141011 & 56941.8 & 7.6 & L & 1499.2 & 800 & 512 & 10.24 & 242.3727(3) & 4.2 \\[0.1cm]
\hline 
\end{tabular} 
\end{center}
\end{table*}

In 2009, observations of \T\ with the 100-m Robert C.~Byrd Green Bank Telescope (GBT) revealed sequences of unusually bright pulses, 
with energies as large as 40 times the mean pulse energy and widths comparable to the width of the average profile. 
Intriguingly, these bright singe pulses (BSPs) were detected predominantly in the vicinity of the eclipses \citep{Bilous2011}. It 
is yet unclear whether BSPs are a new, distinct population of pulses, or the result of a propagation effect.

In this work we analyze eight multi-orbit observations of \T\ from the GBT to better determine BSP properties. 
After describing the observing setup and initial data processing (Sec.~\ref{sec:obs}), we identify the orbital 
phase regions with BSPs and ``normal'' pulses, NSPs (Sec.~\ref{sec:overview}). Based on the time-averaged and 
single-pulse properties from BSP and NSP phase regions, in Sec.~\ref{sec:hyp}, we argue that BSPs are not caused 
by scintillation in the interstellar medium and that they are unlikely to be giant pulses or a special mode of 
emission. Finally, we argue that BSPs, similarly to the unusually strong pulses from PSR~B1957$+$10, 
may in fact be NSPs amplified by lensing on the irregularities in intrabinary material \citep{Main2018}.
In Sec.~\ref{sec:lens} we provide rough estimates of the locations and sizes of the lenses based on the 1D models from \citet{Cordes2017}
and \citet{Main2018}. 

\section{Observing setup and data processing}
\label{sec:obs}

\begin{figure*}
\centering 
\includegraphics[scale=0.95]{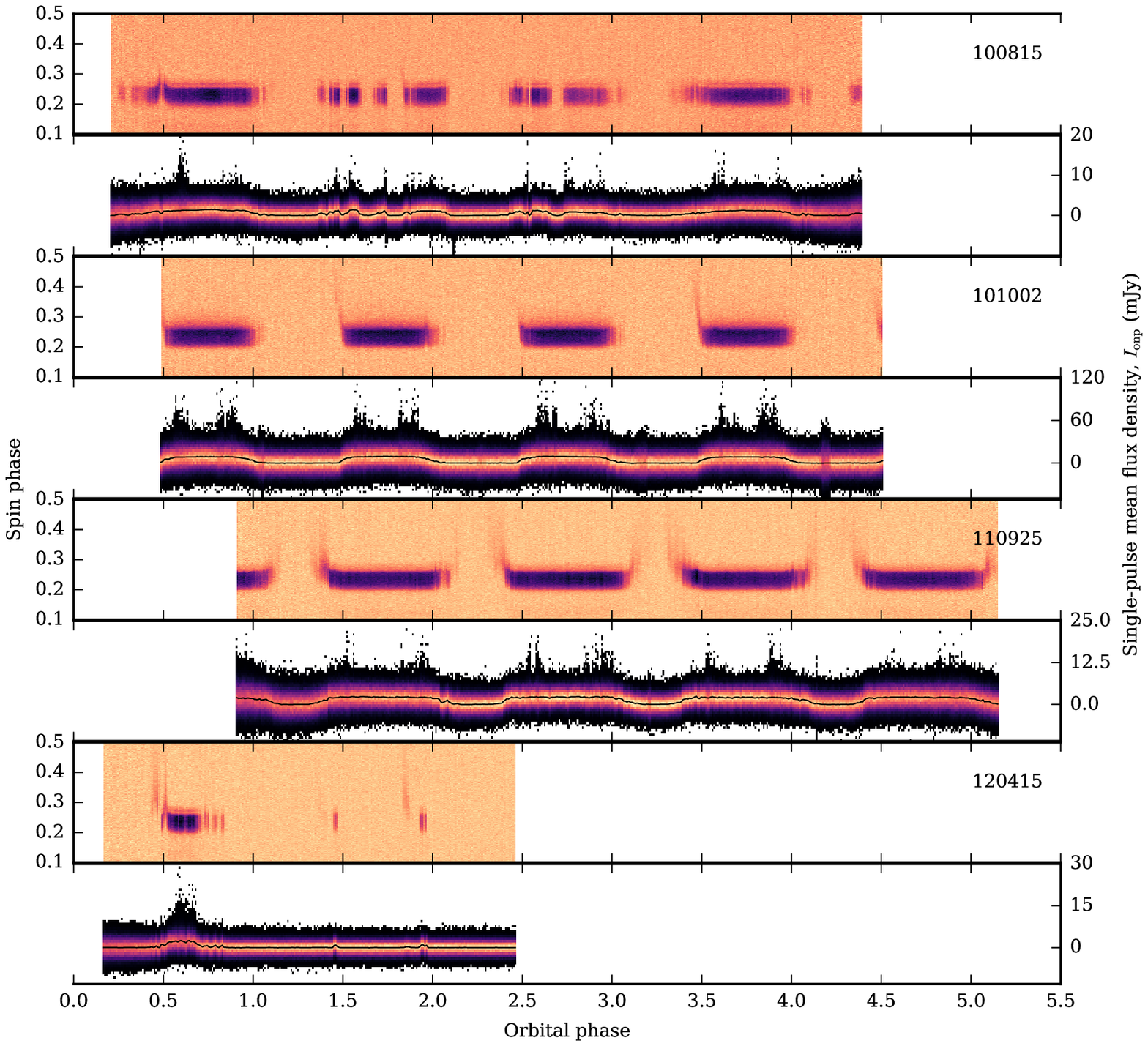}
\caption{Total intensity of the average profile and single-pulse flux density distribution for the first four observing 
sessions, labeled with observation date in \textit{yymmdd} format. For each session, the \textit{upper subplot} shows the 
total intensity of pulsed emission (color) versus pulsar spin and orbital phases integrated over 60\,s. Color scale varies 
from session to session. The sessions are aligned by orbital phase, counted from the first passage of the ascending node.
\textit{Lower subplot}: distribution of the single pulse flux densities (Eq.~\ref{eq:Ionp}) per 5000 pulses (57.5\,s). 
The color indicates the number of pulses in each flux density bin (lighter color corresponds to the larger number of pulses). 
Thin black line shows the average flux density. RFI is visible as an excess of pulses both with larger and smaller flux 
densities (e.g. close to $\phiB \approx 4.2$ for session 101002). The y-axis scale for  pulse energy distribution was clipped, the brightest 
single pulses have flux densities about 2 times larger than the plot limits. }
\label{fig:obs_1} 
\end{figure*} 

Since 2009, the GBT has observed Terzan~5 roughly quarterly, in order to time the known pulsars and search for new ones
\citep{Ransom2005,Hessels2006,Prager2017}.
A typical observing session spans 7--8 hours, or several full \T\ orbits \citep[$P_\mathrm{orb}=1.8$\,hr,][]{Lyne1990}. The 
signal is recorded in one of three radio bands: UHF (central frequency of 820\,MHz), L-band (1500\,MHz), or S-band 
(2000\,MHz)\footnote{The names of the bands follow the standard Institute of Electrical and Electronics Engineers (IEEE) 
radio frequency naming convention.}. 

For this study we hand-picked eight observing sessions with good signal-to-noise (S/N) and diverse eclipse behavior.  
Table~\ref{table:obssum} summarizes the details of these observations. 

The data were recorded with the GUPPI\footnote{\url{https://safe.nrao.edu/wiki/bin/view/CICADA/NGNPP}} pulsar backend in the 
coherent dedispersion search mode \citep{DuPlain2008}. The average dispersion measure (DM) 
of the cluster\footnote{DM = 238.0\,pc\,cm$^{-3}$; \url{http://www.naic.edu/$\sim$pfreire/GCpsr.html}} was used for 
dedispersing the signal within each of 512 channels. As a part of the automated data processing pipeline, the raw filterbank data 
were folded modulo the predicted pulse spin period and integrated every 60\,s with the \verb fold_psrfits  routine from 
the \verb psrfits_utils  package\footnote{\url{https://github.com/demorest/psrfits\_utils}}.  The folded archives contain 
four Stokes parameters, 512 channels, and 512 spin phase bins (22.6\,$\mu$s per bin, more coarse than the raw time 
series resolution of 10.24\,$\mu$s).

Initially, the ephemerides for folding were obtained with the pulse times of arrival (TOAs) away from eclipse regions in the larger subsample of GUPPI 
data spanning years 2008--2017. TOAs were calculated with a single profile template per frequency band, constructed from the average uneclipsed profile 
from a high S/N session and were used to fit for the spin frequency, its derivative, projected semi-major axis, the epoch of the ascending node, and the 
dispersion measure.  Similar to other black widow/redback systems, \T\ exhibits quite large timing irregularities \citep{Nice2000,Bilous2011}, and we 
therefore updated the orbital parameters (namely, projected semi-major axis and the epoch of ascending node) on a per-session basis. 
Because of that, observing sessions are not phase-connected and the fiducial phase is different for each session, resulting 
in different on-pulse window in each session.
For convenience, we added an extra phase offset to shift the on-pulse window to spin phases 0.1--0.3.

While updating the orbital parameters, we also fit for the average dispersion measure over the session (see Table~\ref{table:obssum}). 
Those DMs were determined at orbital phases where the pulsar signal showed no obvious (and often variable) additional dispersive delay from the eclipsing medium.

The folded archives were polarization and flux calibrated using standard techniques, including folded pulsed calibration diode 
measurements at the position of \T\ as well as on and off of standard unpolarized flux calibrators (quasars B1442$+$101 for L-band and 
3C190 for S-band and the UHF observations).  Here, we focus only on the total intensity signal, deferring polarization study to the future work.

To enable long-term archival storage, the raw data from the Terzan~5 observations have been saved with several times lower time and 
frequency resolution and no polarization information (so-called ``subbanded data''). These data were used to extract single-pulse 
spectra with \texttt{dspsr} software\footnote{\url{http://dspsr.sourceforge.net/}, see also \citet{vanStraten2010}}. The spectra were 
produced for each pulsar rotation, regardless of the presence of signal in the on-pulse region, and consisted of 282 phase bins (spanning 
a whole rotation period) and either 64 or 128 channels (i.e.,~subbands) per band.

Initially, single-pulse data were calibrated with the same procedure as the folded data. However, later it appeared that no scales/offsets 
were applied to the raw data during the single-pulse extraction process.  This resulted in the single-pulse spectra (integrated within 
roughly one minute) differing from the spectra of the folded data: the spectral index of the former was much steeper and the standard deviation 
in the off-pulse region did not increase as \T\ approached horizon. Since the S/N of the pulsed emission was mostly the same for folded and 
single-pulse data (except for the UHF session 101002, for which the folded data had 20\% larger S/N), the single-pulse data archives were 
renormalized to match the system equivalent flux density (SEFD) of the folded data. This was done in the following way: first, the standard 
deviation in the off-pulse region, $\sigma_\mathrm{off}$, was computed for both folded and singe-pulse data in each of the 60-s subintegrations 
and in 32 subbands.  Then, the $\sigma_\mathrm{off}(t)$ sequences were filtered with an 11 minute-long median filter to reduce the influence of 
radio frequency interference (RFI). After that, a smooth spline was fit to the filtered $\sigma_\mathrm{off}(t)$ and single-pulse spectra were 
scaled by the ratio of spline values. An additional coefficient of 1.2 was applied to the UHF session. Resulting band-integrated total intensities 
of single-pulse data, integrated within 60\,s, matched well the corresponding intensities of the folded data.

Since RFI excision of folded/single-pulse data was performed on archives with different time/spectral resolution, such SEFD scaling may lead 
to some distortion of the calibrated spectral  shape of single pulses. However, in this work we examine only the band-integrated total intensity 
of the single pulses or the frequency-resolved S/N. 

\begin{figure*}
\centering 
\includegraphics[scale=0.95]{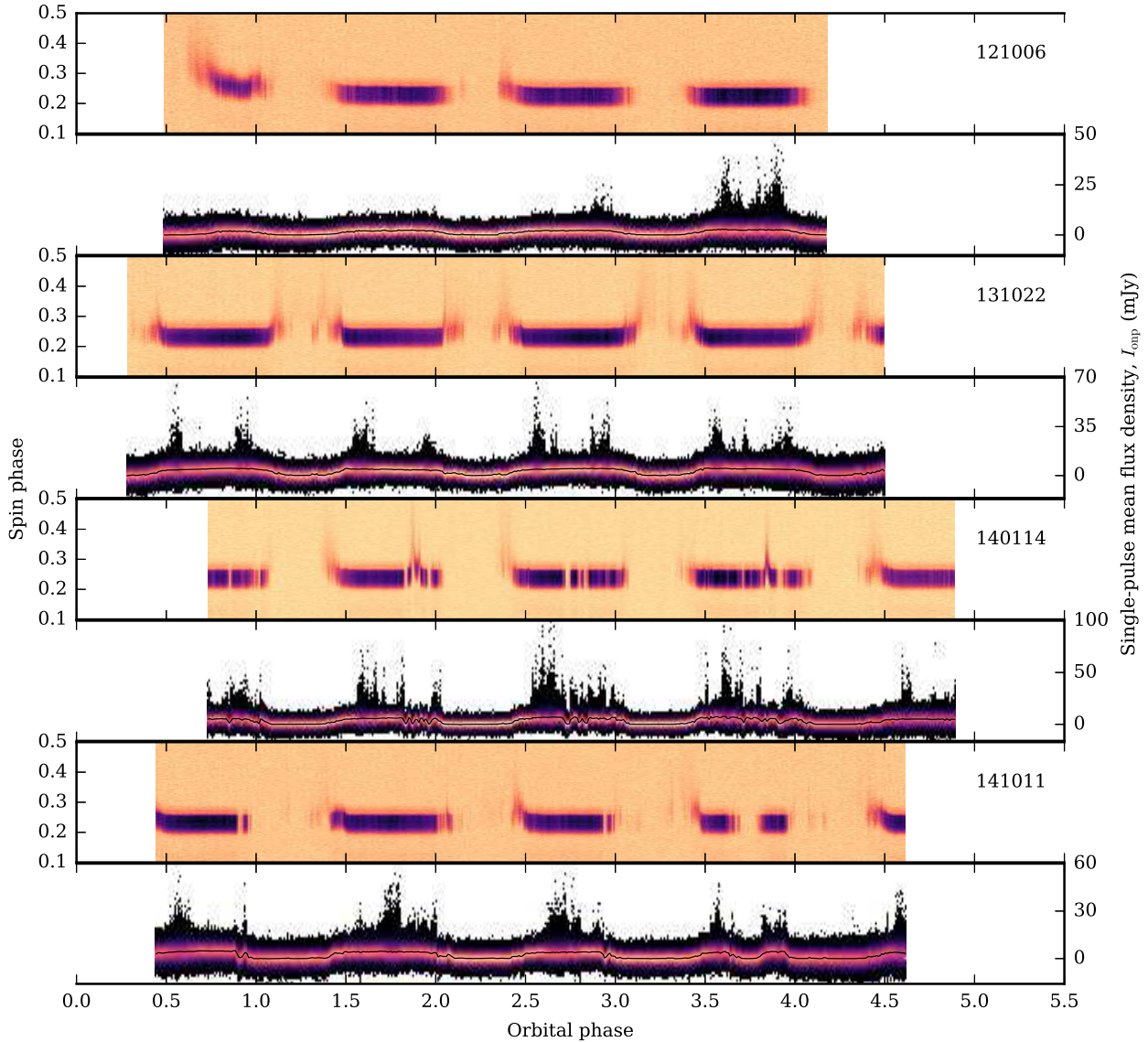}
\caption{Total intensity of the average profile and single-pulse flux density distribution for the last four observing sessions. See the 
caption of Fig.~\ref{fig:obs_1} for more details.}
\label{fig:obs_2} 
\end{figure*} 

\section{Pulsar behavior during observing sessions}
\label{sec:overview}

Figures~\ref{fig:obs_1} and \ref{fig:obs_2} provide an overview of the average emission and the single pulse flux density distributions for all 
observing sessions. For each pair of subplots, the upper subplot shows total intensity vs. spin and orbital phases for the 60-s data folds. 
The lower subplot shows the distribution of the single-pulse flux densities, which  were calculated as the sum of intensity samples in the 
on-pulse window (spin phase 0.18--0.3) divided by the number of bins in one period:
\begin{equation}
\label{eq:Ionp}
I_\mathrm{onp} = \frac{1}{N_\mathrm{bin}}\sum^{\phi=0.3}_{\phi=0.18} I(\phi).
\end{equation}
The distributions were calculated for every 5000 periods or 57.7\,s.

In all sessions \T\ is eclipsed around the inferior conjunction ($\phiB = 0.25$), with eclipse typically lasting almost a half of an orbit. 
Near the ingress/egress the intensity of the pulsed emission is lower and dispersive tails are usually visible (e.g.,~session 110925). Sometimes 
a faint dispersed signal is present throughout most of the normally eclipsed orbital phase range (e.g.,~session 131022). Some of the sessions 
exhibit short eclipses at random orbital phases, with or without dispersive tails (e.g.,~session 140114). During some of the sessions the signal 
is present only in a small fraction of an orbit (e.g.,~session 120415). All of this behavior has been previously described in the literature 
\citep{Thorsett1991,Nice1992}. During one of our sessions, namely 121006, we witnessed the slow dissipation of an intrabinary plasma cloud 
which was enshrouding the pulsar prior to the session start: the dispersion measure outside of normal eclipse orbital phases gradually 
decreased for four consecutive orbits.

Because the on-pulse window was rather small and fixed in spin longitude, the flux density distributions are less accurate in the regions with 
larger dispersion. During eclipses, when no pulsed signal is present, the flux density distributions match the noise distributions. Sometimes, 
despite the zapping, RFI can be seen as the excess of both high and low energy flux density values (e.g.,~at $\phiB\approx4.25$ of session 101002). 

It is evident that the distribution of single-pulse flux densities varies considerably in shape, with bursts of bright single pulses occurring 
close to the ingress and egress of the eclipses. Sometimes BSPs are detected throughout all uneclipsed $\phiB$ (e.g.,~140114/3, with ``/3'' 
indicating the third orbit, with $\phiB$ between 2 and 3). 

\subsection{Defining orbital phase regions with BSPs}

By definition, BSPs are pulses with unusually large flux densities. However, defining orbital phases with BSPs based on 
the presence of pulses with flux densities above the certain threshold makes use of the tail of the single-pulse flux density 
distribution and thus is prone to statistical fluctuations due to the  small number of pulses in the tail. Thus, instead of setting 
a threshold for the flux density of BSPs, we examine the standard deviation of the flux densities for all pulses in blocks of data of 
a certain length. The procedure for that is described in Appendix~A and the orbital phase limits for RFI, BSP, and NSP $\phiB$ regions 
for all sessions are shown in Figs.~\ref{fig:sps_stat_1}--\ref{fig:sps_stat_4}(b). 

Typical durations where BSPs are prominent were estimated from the total length of the consecutive BSP blocks. These estimates obviously 
depend on the choice of the standard deviation threshold and the size of the block. Nevertheless, it can be said that bursts of strong 
pulses last approximately 1--15 min, with the durations varying between sessions. BSPs can be present for as long as 20\,min (e.g.,~140114/3) 
and as short as 10\,s (131022/1, computed with smaller block size). 
Some sessions exhibit dense sequences of distinct bursts which last for tens of seconds.

\section{What are BSPs?}
\label{sec:hyp}

Having separated the orbital phases with BSPs and NSPs, we can examine the differences in pulsar emission properties and the intra-binary 
environment with the ultimate goal of constraining the possible nature of BSPs.

\begin{figure}
\centering 
\includegraphics[scale=1.0]{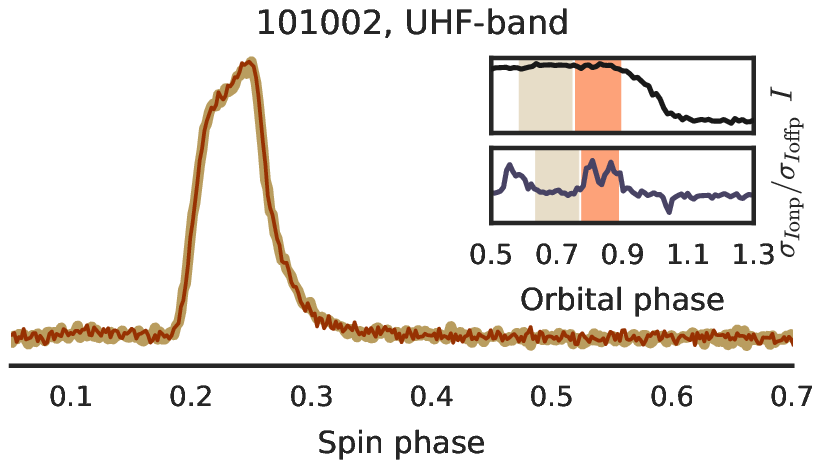}
\includegraphics[scale=1.0]{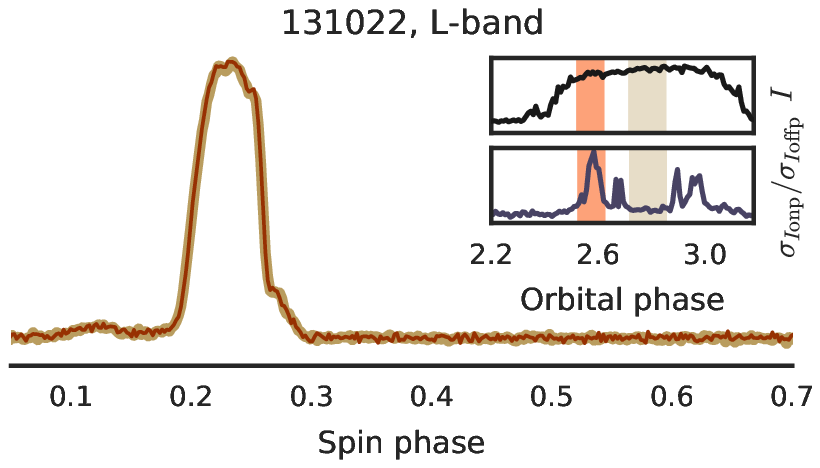}
\includegraphics[scale=1.0]{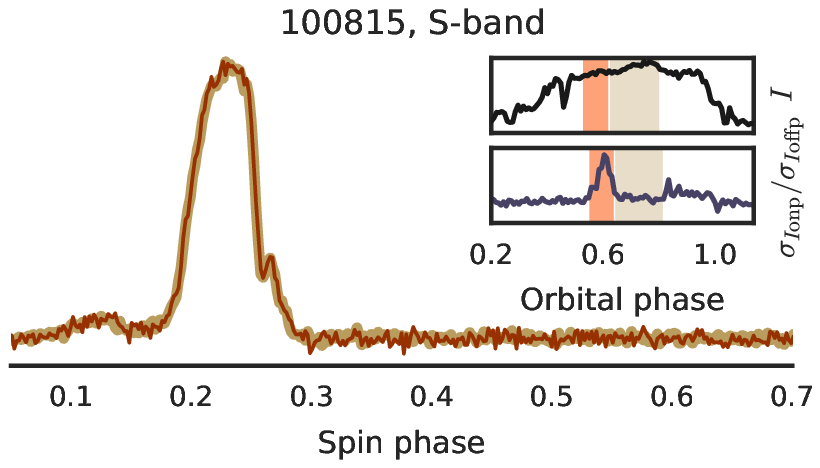}
\caption{Average pulse  profile in three frequency bands in the BSP and NSP $\phiB$ regions (brown and beige lines, respectively). Profiles
were normalized by their peak intensities. Insets show the average flux density (\textit{top}) and the ratio of standard deviations of flux density 
values in the on- and off-pulse regions (\textit{bottom}) versus orbital phase. The corresponding BSP and NSP regions are highlighted with brown 
and beige colors (cf.,~Figs.~\ref{fig:sps_stat_1}--\ref{fig:sps_stat_4}(a) and see text for more details).}
\label{fig:AP} 
\end{figure}

\subsection{Scintillation on the ISM?}

Because of scintillation on the inhomogeneities in the ionized ISM, observed pulsar emission is always modulated in radio frequency and time. 
This modulation manifests itself as the regions of enhanced flux density (called scintles) in the dynamic spectrum. The size of the scintle 
in frequency (also called the decorrelation bandwidth) is set by the electron density variations in the ISM and the temporal scale of the 
modulation depends on the velocities of the pulsar and ISM plasma relative to observer 
\citep[see][Chapter 4, for the review]{Lorimer2005}.  

In the strong scintillation regime, when decorrelation bandwidth and scintillation timescale are comparable to the observing 
bandwidth and integration time, respectively, the observed pulsar flux appears to be strongly modulated. 
No scintles, however, were found upon visual examination of \T's dynamic spectra, averaged over 10--60\,s and retaining the original frequency 
resolution of 0.4\,MHz in UHF and 1.6\,MHz in L- and S-bands. Moreover, the decorrelation bandwidth, estimated as the inverse of the scattering 
timescale, appeared to be much smaller than the channel width in all of the three bands.  The scattering timescale was measured 
by fitting a model of \T's average profile, convolved with 
a thin-screen scattering kernel, to the time-integrated UHF data away from eclipses. To do this, we used a routine from the Pulse 
Portraiture code \citep{Pennucci2014,Pennucci2016}, which 
fits a 2D (phase/frequency) model to the data, allowing to account for the profile evolution.
We have modelled the average profile with 3--4 Gaussian components (the results did not depend much on the exact number of components used) with 
frequency-independent locations and widths, but frequency-dependent amplitudes. 
The fits yielded scattering timescales of about 210\,$\mu$s 
(0.018 of spin phase) at 820\,MHz, with the scattering index of about $-3.3$. This indicates a decorrelation bandwidth of 0.8\,kHz, much smaller 
than the width of one frequency channel. At the highest frequency in our observing setup, 2400 MHz, the decorrelation bandwidth scaled with either 
measured or Kolmogorov power-law indices is still much smaller than the channel width. 
That means that we are averaging over many scintles and the scintillation on the ISM is not the cause of the observed bursts of pulses.

It is worth noting that the shape of the average profile and the measured scattering timescales are the same in the both BSP and NSP regions, 
further strengthening the conclusion.

\subsection{Mode changes?}

Some pulsars switch between two quasi-stable radio emission modes, which differ in the shape of the radio profiles and (for at least some PSRs) 
in the properties of the single pulses. For some of these pulsars, modes have been linked to changes in the spin-down rate \citep{Lyne2010} 
and correlated variation of the X-ray emission \citep{Hermsen2013}. This suggests that radio mode switching is an indicator of some kind of 
global magnetospheric transformation, although the details are far from being fully understood and it is unclear whether all observed 
mode-switching phenomena are caused by the same process.

A few relatively young isolated pulsars exhibit so-called ``bursting modes'', namely, PSRs B0611+22 \citep{Seymour2014}, J1752+2359 \citep{Gajjar2014}, 
and J1938+2213 \citep{Lorimer2013}. Bursting modes are characterized by an abrupt onset, higher average intensity, relatively large spread of 
single-pulse energies, and changes in the shape of the average profile. At least for PSRs B0611+22 and J1752+2359, bursting modes occur in a 
quasi-periodic fashion and  at least for PSR B0611+22, the relative flux density between the modes depends on radio frequency, indicating a 
changing radio spectral index \citep{Rajwade2016}. 

Rapid onset and higher $\sigma_{I_\mathrm{onp}}$ make BSP regions similar to the bursting modes. However, the shape of the average profile is 
the same for BSPs and NSPs (Fig.~\ref{fig:AP}) and neither band-integrated flux density nor the intra-band spectral index of radio emission 
exhibit clear change between BSP and NSP regions (Figs~\ref{fig:sps_stat_1}--\ref{fig:sps_stat_4}, a, e, f).

\begin{figure}
\centering 
\includegraphics[scale=0.9]{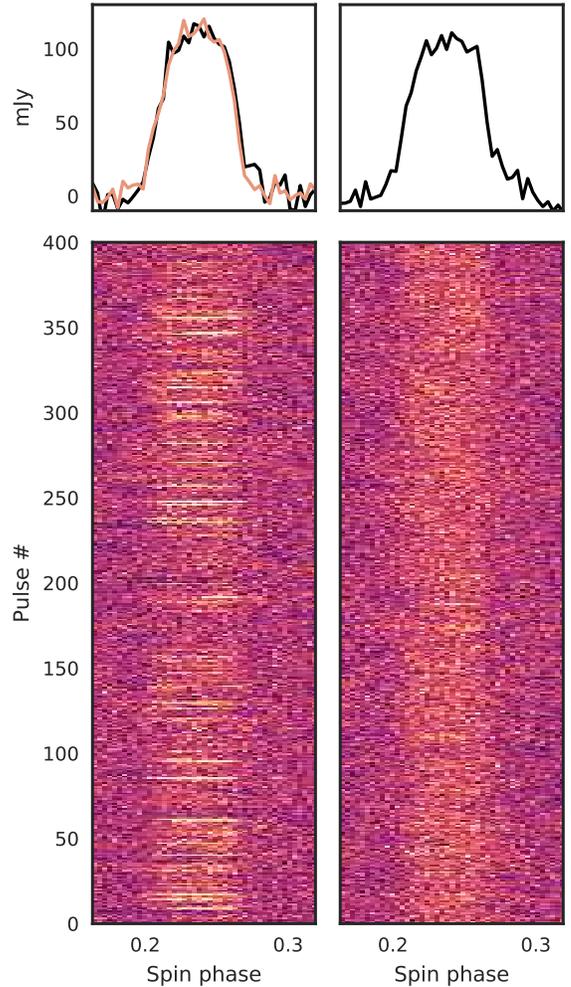}
\caption{An example of pulse sequences from adjacent BSP (left) and NSP (right) regions
(session 140114/2, $\phiB$ of 1.74 and 1.61, respectively). In order to show the fainter pulses better, the colormap is saturated at $\pm4$ 
standard deviations in the off-pulse region. The peak S/N values in the BSP and NSP $\phiB$ regions are 11.8 and 5.0, respectively. The top 
panels show the average profiles of each 400-pulse sequence (black), plotted on the same flux density scale. The light line shows the average 
of pulses with peak $\mathrm{S/N}>5$, with intensity scaled down by a factor of 5.}
\label{fig:sps_sequence} 
\end{figure} 

\begin{figure*}
\centering 
\includegraphics[scale=0.66]{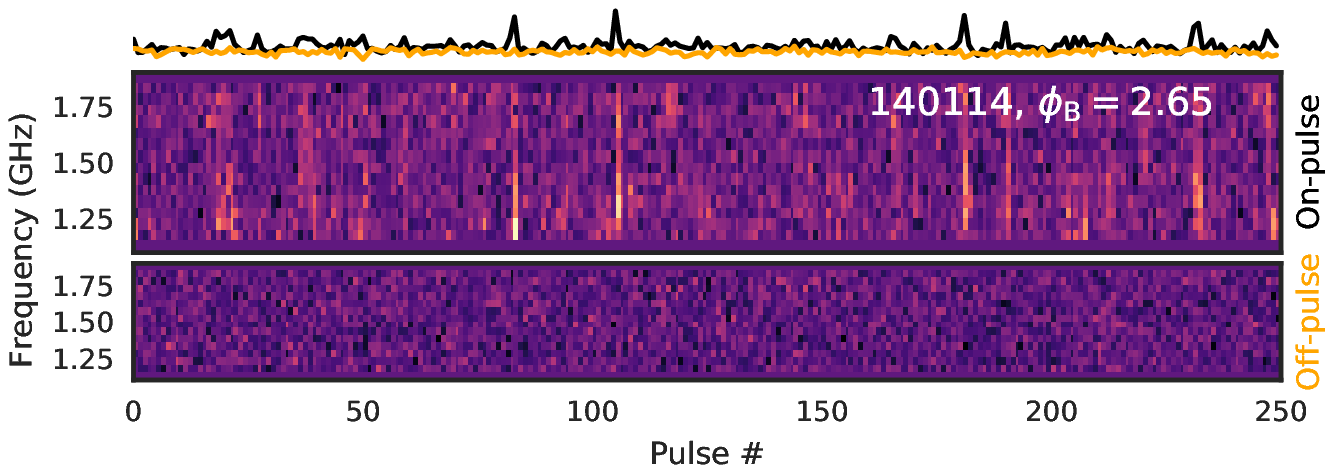}\includegraphics[scale=0.66]{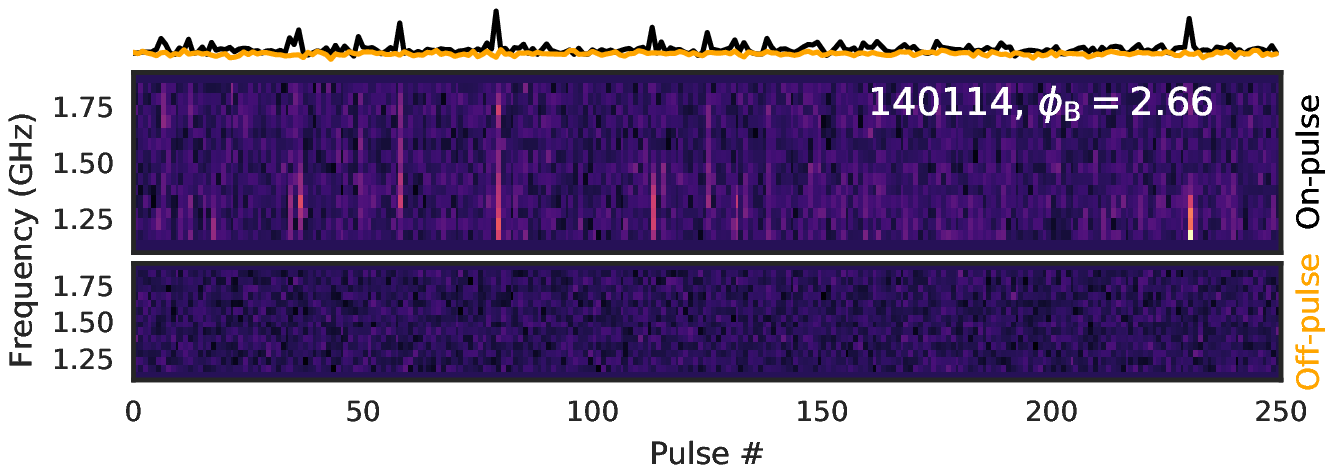}
\includegraphics[scale=0.66]{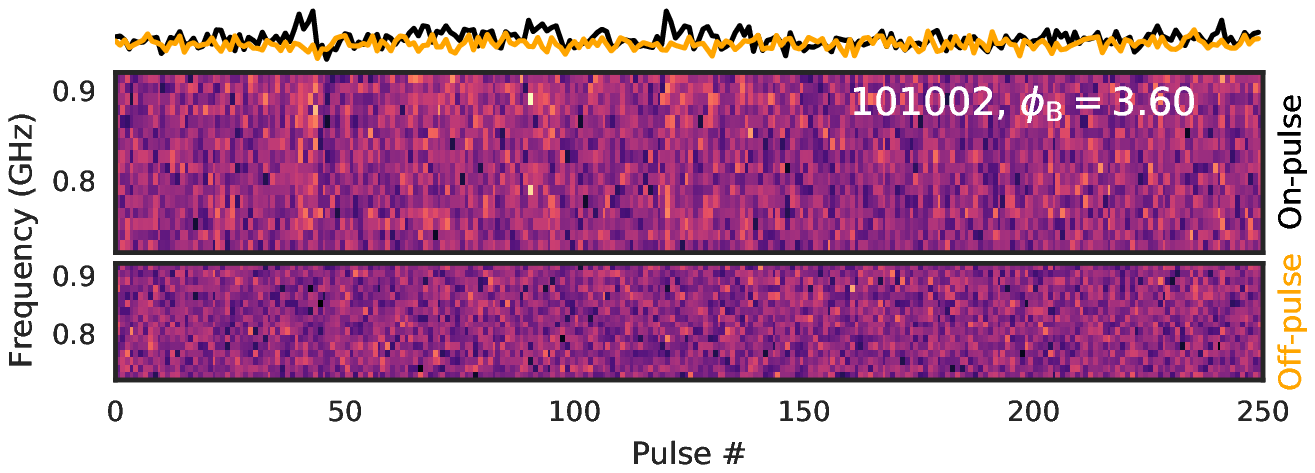}\includegraphics[scale=0.66]{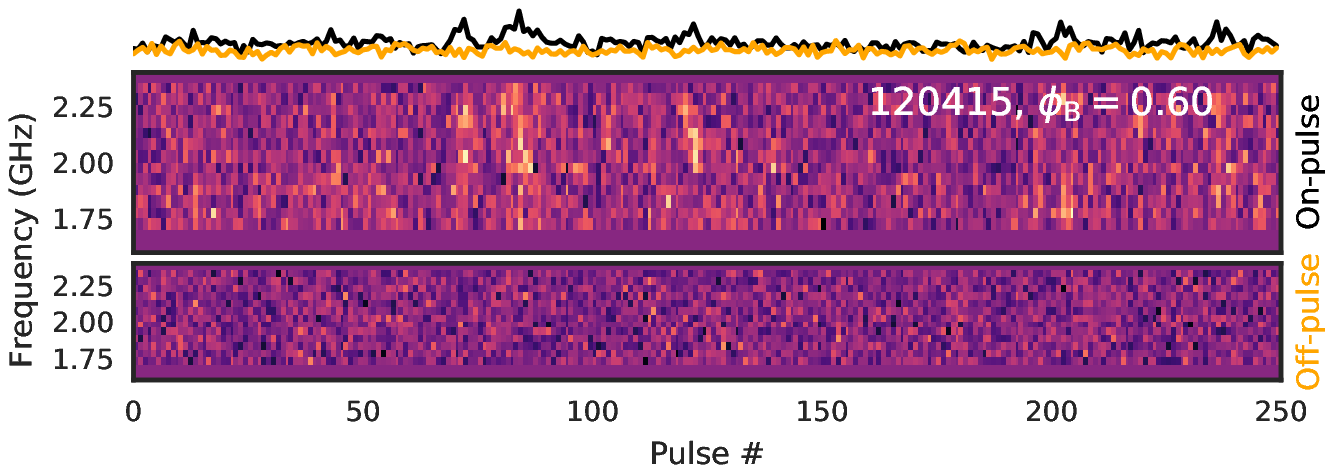}
\includegraphics[scale=0.66]{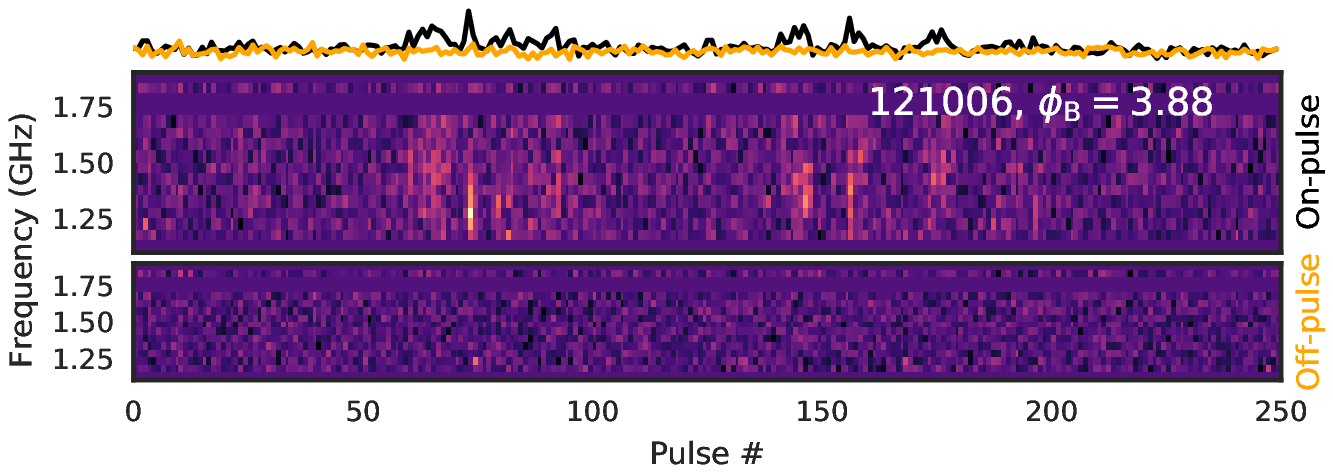}\includegraphics[scale=0.66]{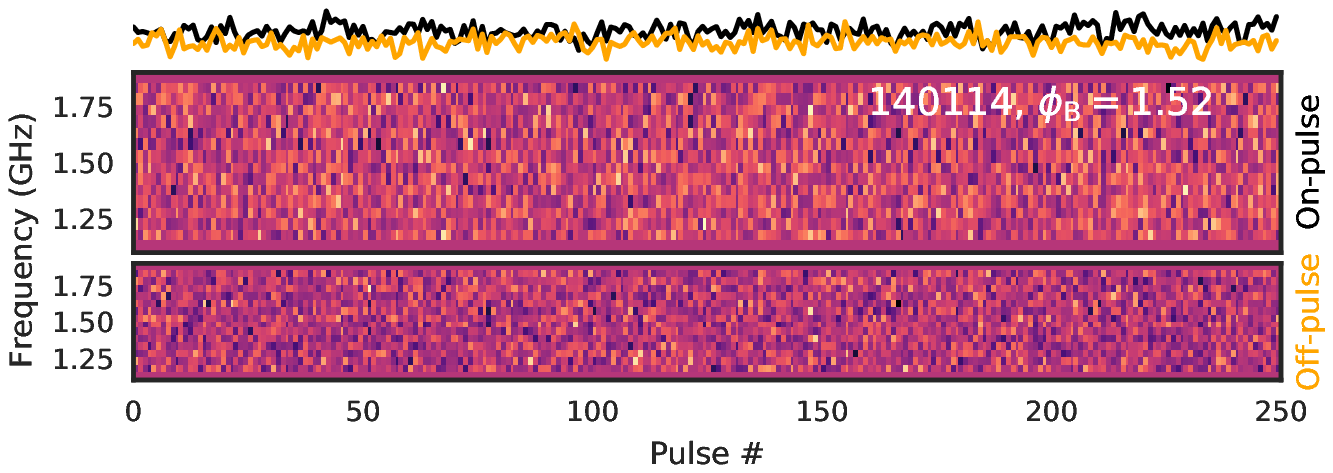}
\caption{An example of dynamic spectra of several selected pulse sequences. Band-integrated S/N in the  on-pulse (black) and off-pulse 
(orange) windows is shown on top of each panel, with dynamic spectra integrated within on-pulse (middle subplot) and off-pulse (bottom) windows. 
Both on- and off-pulse dynamic spectra are plotted with the same color scale where lighter colors correspond to larger S/N. \textit{Top row}: an 
example of some of the strongest BSPs at L-band. \textit{Middle row} and \textit{bottom row, left}: BSPs in UHF, L-, and S-bands showing correlated 
structure in frequency and time. \textit{Bottom row, right}: sample dynamic spectra of NSPs.}
\label{fig:dynspec} 
\end{figure*} 

\subsection{Giant pulses?}

Giant pulses (GPs) are rare and mysterious radio pulses that have been detected only from a handful of pulsars, all of them either young or recycled. 
GPs are characterized by their short durations (ns--$\mu$s), large brightness temperatures \citep[up to $2\times10^{41}$\,K,][]{Hankins2007}, and 
power-law energy distribution \citep[see review by][]{Knight2006a}. GPs usually come from the narrow spin longitude regions on the trailing or leading edge of the average profile.
The nature of GPs is unclear, but they are thought to originate close to the light cylinder by a separate  emission mechanism \citep{Hankins2016}. 

The typical duration of a BSP ($\tau \approx 0.8$\,ms) is much longer than that of GPs and is comparable to the width of \T's average pulse profile 
(Fig.~\ref{fig:sps_sequence}). The energy of BSPs, $E=IP/\tau$ may occasionally exceed the formal threshold for GPs, namely $E>20\langle E \rangle$, 
where $\langle E \rangle$ is the average pulse energy.
Assuming that the size of the emitting region is equal to $c\tau$, the brightness temperature of a strong BSP can be calculated as follows \citep{Soglasnov2004}:
\begin{equation}
T_\mathrm{b}= \frac{Id^2P}{k\nu^2\tau^3}.
\end{equation}
Here $I$ is mean pulse flux density (taken to be 140\,mJy, the brightest pulse from session 141011), 
$k=1.38\times10^{-16}$\,erg K$^{-1}$ is Boltzman's constant, $d$ is the distance to \T\ \citep[5.9\,kpc,][]{Valenti2007}, and 
$\nu=1500$\,MHz is the observing frequency. These parameters result in $T_\mathrm{b} \sim 10^{25}$\,K, much smaller than $10^{35}$--$10^{41}$\,K 
brightness temperature of ``classical'' GPs from PSR B1937+21 \citep{Soglasnov2004} or the Crab pulsar \citep{Hankins2007}. 

The energy distribution of BSPs may be described as power-law if only the tail is considered. Roughly, the power-law index ranges between $-3$ and 
$-5$ for the cumulative density distribution, which is steeper than the corresponding power-law index for GPs \citep[ranging from $-1$ to $-3$,][]{Knight2006a}.

\subsection{Lensing on irregularities in the ablated material?}

Before reaching the observer, \T's radiation travels through highly dynamical and irregular streams of plasma which are ablated from the surface of 
companion star. These plasma streams are responsible for observed DM fluctuations, they affect the spectral index of pulsar radio emission, and, 
ultimately, quench pulsed emission during eclipses. Below we will argue that BSPs may result from strong lensing on the irregularities in the ablated plasma. 

The incident wavefront of pulsar radio emission is focused (or defocused) on plasma irregularities, and, under favorable circumstances, this 
process may take place in a caustic regime, resulting in strong (by a factor of 10--100) amplification of individual radio pulses. Lensing on DM 
irregularities has been proposed as an explanation for unusually strong pulses from the original Black Widow pulsar B1957+10\footnote{Giant pulses 
have also been detected from this pulsar \citep{Joshi2004,Knight2005}.}. These unusually strong pulses share similar properties with \T's BSPs\footnote{Namely, 
coming in groups in the orbital phases close to eclipses, having widths comparable to the width of the average pulse, and displaying chromatic spectra.} 
\citep{Main2018}. \citet{Cordes2017} speculated that plasma lensing may be responsible for some of the Fast Radio Bursts, bright solitary radio pulses of unclear origin. 

Strong lensing effectively leads to the redistribution of signal power in frequency and time, allowing for a natural 
explanation of the following phenomena: (a) no apparent change in the average flux density during BSP outbursts; (b) unchanging shape of the average 
profile in the adjacent BSP and NSP orbital phase regions (also true for profiles made from very bright pulses only); (c) similar spectral index of 
average emission in BPS and NSP regions, and (d) in addition to the presence of unusually bright pulses, a larger number of weak pulses in BSP orbital 
phase regions as compared to NSPs (see Appendix~\ref{app:efit}).

The maximum observed amplification of single-pulse flux density, defined as the ratio of flux density values of the 
strongest BSPs to the strongest NSPs (Appendix~\ref{app:efit}) is $G\approx 10$, which is typical for lensing \citep{Cordes2017,Main2018}. 
Dynamic spectra of BSPs show structures correlated in frequency/time on the scale of about 200\,MHz and $\sim 50$\,ms 
($\sim 5$ spin periods), respectively (Fig.~\ref{fig:dynspec}). This is at odds with the featureless dynamic spectra of NSPs (although the lack of 
features can be influenced by the relative weakness of the individual NSPs). 
Unlike \citet{Main2018}, we do not observe clear slanted structures (or ``slopes'') in our dynamic spectra, although some hints of those may be present. 
It is possible that there are multiple lenses present, producing an interference pattern of the lensing images and destroying obvious slopes.  

\begin{figure}
\centering 
\includegraphics[scale=1.0]{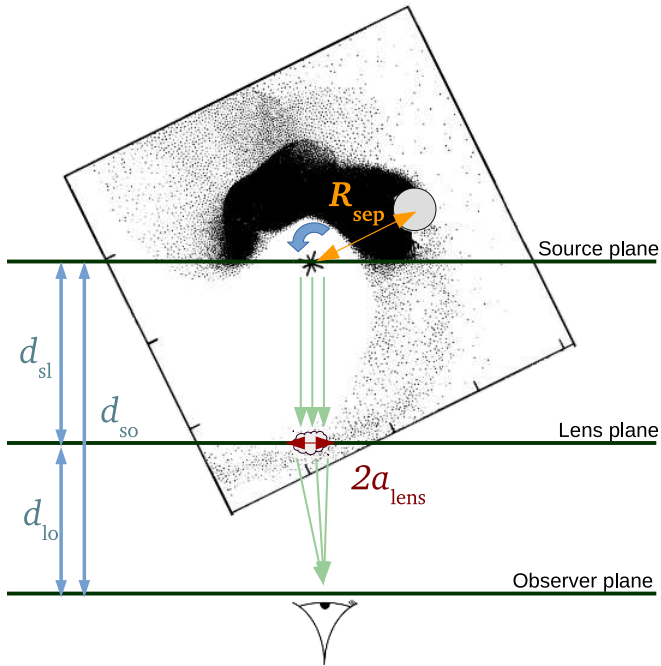}
\caption{Schematic depiction of lensing configuration. The background shows a hydrodynamical 
model of plasma outflow from \citet{Tavani1991}. The pulsar is marked with a star in the center of the plot and the Roche 
lobe of the companion is shown with a gray circle. Counterclockwise direction of orbital motion is shown with the blue arrow. The locations of pulsar, 
companion and observer correspond to $\phiB\approx0.6$, one of the two  orbital phase regions where BSPs are most likely to occur 
(cf.,~Figs.~\ref{fig:sps_stat_1}--\ref{fig:sps_stat_4}, middle panels).} 
\label{fig:lens_sketch} 
\end{figure}

\T's BSPs tend to cluster in two orbital phase windows around $\phiB\approx0.6$ and 0.9 (Figs.~\ref{fig:sps_stat_1}--\ref{fig:sps_stat_4}, middle panels), 
although occasionally BSPs occur throughout most of the uneclipsed orbital phase regions (e.g.,~session 140114). This clustering may be explained with 
the shape of the plasma outflow. Fig.~\ref{fig:lens_sketch} shows an example of an outflow configuration from 2D hydrodynamical simulations\footnote{The model 
assumed the gaseous material to be optically thick to pulsar radiation everywhere except for the thin layer directly exposed to the radiation. The calculations 
included bremsstrahlung cooling, gravitational force, gas pressure, centrifugal and coriolis forces. The authors obtained results for several values of the mass 
loss rate, for Mach number of the order of 1, for injection velocities on the order of escape velocity from the companion star, and pressure balance.} in the works 
of \citet{Tavani1991,Tavani1993}. The outflow (shown in the corotating reference frame) has a relatively less dense tail which bends around and affects pulsar 
emission at larger spans of orbital phases, roughly similar to the orbital phases where BSPs are observed (cf.,~Fig.~\ref{fig:lens_sketch} and the middle 
panels of Figs.~\ref{fig:sps_stat_1}--\ref{fig:sps_stat_4}). However, the simulation of \citeauthor{Tavani1991} contains a single tail, whereas BSPs occur 
symmetrically around eclipses, requiring a more symmetrical outflow configuration. It is worth mentioning that another model of the plasma outflow in \T\ system, 
the optically thin model of \citet{Rasio1991} yielded two symmetric tails. 

\begin{table}
\begin{center} 
\caption{Binary system parameters\label{table:syspar}}
\begin{tabular}{ll} 
\hline 
Spin period & $P = 11.56$\,ms \\
Orbital period & $P_\mathrm{B} = 6536$\,s \\
Distance to source plane & $d_\mathrm{so} = 5.9$\,kpc \\
Distance to lens plane & $d_\mathrm{lo} = 5.9$\,kpc \\
Projected semimajor axis & $A\sin i = 0.12$\,lt-s \\
Separation between PSR and companion\footnote{In the absence of rigorous constraints, assuming $i=45$.} & $R_\mathrm{sep} = 2.5$\,lt-s\\
Derived orbital velocity of companion & 680 km/s\\
\hline
\end{tabular} 
\end{center}
\end{table}

\section{Estimates of the size and location of plasma lens}
\label{sec:lens}

The size and location of the lens may be roughly estimated following frameworks established in \citet{Cordes2017} and \citet{Main2018}.
\citet{Cordes2017} examined strong lensing from a single 1D lens of characteristic size $a_\mathrm{lens}$. 
DM within the lens was chosen to have Gaussian distribution with the maximum DM excess of $\mathrm{DM}_l$.
We will parametrize the size and location of the lens with dimensionless quantities $a$ and $d$, defining the size of 
the lens as $a_\mathrm{lens} = a\times R_\mathrm{sep}$, and the distance from the pulsar to the lens as 
$d_\mathrm{sl} = d\times R_\mathrm{sep}$, where $R_\mathrm{sep}$ is the separation between pulsar and companion. 
Table~\ref{table:syspar} lists $R_\mathrm{sep}$ as well as several other system parameters. 

Maximum pulse amplification $G$ is set by the Fresnel scale and the size of the lens. The Fresnel scale at the lens plane (Fig.~\ref{fig:lens_sketch}) is given by:
\begin{equation}
r_\mathrm{F} \approx \sqrt{\frac{c d_\mathrm{sl}d_\mathrm{lo}}{\nu d_\mathrm{so}}} \approx 12.4\times \sqrt{d}\,\,\mathrm{km}
\end{equation}
for $\nu=1.5$\,GHz. According to \citet{Cordes2017}, who obtained amplification $G$ by evaluating the Kirchoff diffraction integral,
\begin{equation}
\label{eq:Gmax}
G \sim \frac{a_\mathrm{lens}}{r_\mathrm{F}} \approx 6.1\times 10^5 \frac{a}{\sqrt{d}}. 
\end{equation}
Since the maximum gain is about 10:
\begin{equation}
\frac{a}{\sqrt{d}} \sim 1.6\times 10^{-4}. 
\end{equation}

Further constraints on $a$ and $d$ can be placed from the time of caustic crossing. Following Eq.~22 in
\citet{Cordes2017}:
\begin{equation}
t_\mathrm{c} \sim \frac{a_\mathrm{lens} (\delta G/G)}{v_\mathrm{trans}G^2} \left(\frac{d_\mathrm{lo}}{d_\mathrm{so}}\right)\approx 11\times a\,\mathrm{s}, 
\end{equation}
where $v_\mathrm{trans}\approx 680$\,km/s is the effective transverse velocity. Since the lens is much closer to the source than to observer, the transverse 
velocity is set by the velocity of the plasma outflow, which is in turn assumed to be close to the companion's orbital velocity \citep{Tavani1991,Tavani1993}. 
The fractional change in gain $\delta G / G $ is taken to be $\approx 10/10 = 1$. 
Setting $t_\mathrm{c} \approx 1$--$50$\,ms, we obtain $a \approx 10^{-4}$--$5\times10^{-3}$ 
(corresponding to $a_\mathrm{lens}\approx68-3400$\,km) and $d \approx 0.3$--$800$.

The presence of lensing sets an upper limit on the size of the emission region, $\delta x$. According to Eq.~18 from \citet{Cordes2017}, 
$\delta x \lesssim 9\times d/a$\,cm, or 0.3--17\,km for the obtained lens parameters. The size of the emission 
region is therefore smaller than the upper limits from the light crossing time ($c\tau\approx 300$\,km), but 
is comparable to the upper limits obtained from the influence of scintillation on flux density statistics 
of the Vela pulsar \citep{Johnson2012} and very long baseline interferometric imaging of the interstellar 
scattering speckle patterns associated with the pulsar PSR B0834+06 \citep{Pen2014}.

\begin{figure}
\centering 
\includegraphics[scale=1.0]{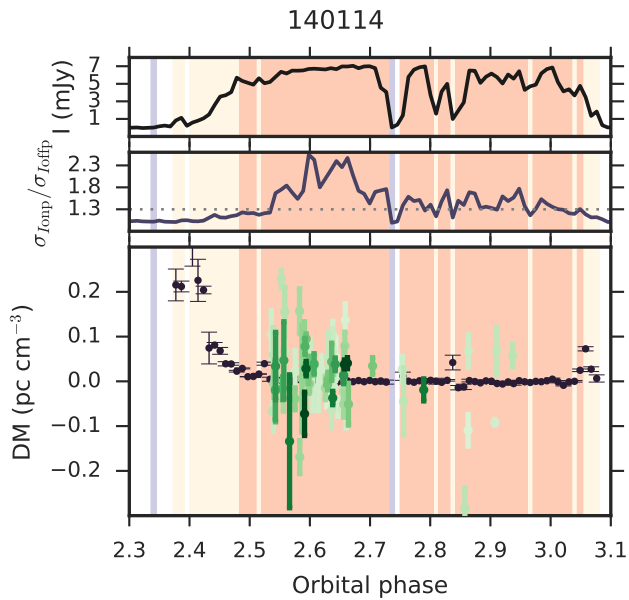}
\caption{ Panels (a), (b) and (c) from Fig.~\ref{fig:sps_stat_4}, for the second orbit from session 140114.  The DMs of individual BSPs are 
overplotted with green errorbars. Darker colors correspond to stronger pulses. The DM of individual BSPs were measured 
with the same method as DMs from the time-integrated emission.}
\label{fig:DM_BSPs} 
\end{figure}

The lower limit on the lens DM, $\mathrm{DM}_l$, can be estimated from the fact that in order for strong lensing to occur, the distance from the lens to 
the observer must be larger than the focal distance and the observing frequency must be smaller than the focal frequency. Using Eqs.~7--8 from \citet{Cordes2017} 
we obtain the focal distance and frequency:
\begin{equation}
d_\mathrm{f} = 8500\times\frac{a^2}{d}\times\left(\frac{\mathrm{DM}_l}{1\, \mathrm{pc}\, \mathrm{cm}^{-3}}\right)^{-1}\,\mathrm{kpc},
\end{equation}
and 
\begin{equation}
 \nu_\mathrm{f} = 0.04\times\sqrt{\frac{\mathrm{DM}_l}{1\, \mathrm{pc}\, \mathrm{cm}^{-3}}}\times\frac{\sqrt{d}}{a}\,\mathrm{GHz}.
\end{equation}
For $a/\sqrt{d}=1.6\times10^{-4}$, to satisfy $d_\mathrm{lo} > d_\mathrm{f}$ and $\nu<\nu_\mathrm{f}$, the lens DM must be larger than
$4\times10^{-5}$\,pc cm$^{-3}$. Unfortunately, existing data do not allow stringent constraints on $\mathrm{DM}_l$: 
the narrow-band frequency structure, large width and moderate (in comparison to the time-averaged profile) S/N of individual BSPs result in large errors 
on single-pulse DM measurements. For our observations, DMs measured with the brightest BSPs had errors on the order of 0.1\,pc/cm$^3$ (Fig.~\ref{fig:DM_BSPs}) 
and individual DMs were scattered around DMs from the 60-s subintegrations by the same amount. Future and much more sensitive multi-frequency 
observations could provide better estimates of $\mathrm{DM}_l$. 

\citet{Main2018} explored lensing on a perfect eliptical lens, 
with a single focal point to which all paths within 
the lens contribute coherently. For an extreme case of one-dimensional lens, the 
gain in intensity scales as the square of the lens size and Fresnel scale ratio: $G = \pi(a_\mathrm{lens}/r_\mathrm{F})^2$. At $\nu=1.5$\,GHz,
this results in
\begin{equation}
\frac{a}{\sqrt{d}} \sim 3\times 10^{-5}. 
\end{equation}
The full-width half-maximum duration of strongly magnified events is calculated as:
\begin{equation}
 t\approx \frac{1.4 r_\mathrm{F}^2}{\pi v_\mathrm{trans}a_\mathrm{lens}} \approx 4\times 10^{-3} \sqrt{d}\,\mathrm{s}.
\end{equation}
Together, these equations yield $d$ of 0.06--160 and $a$ of $7\times10^{-6}-4\times10^{-4}$, corresponding to $a_\mathrm{lens}=5-280$\,km.

\section{Summary}
\label{sec:summary}

Although the general picture of the \T\ binary system was established very soon after its discovery in 1990, the phenomena discovered by radio 
observations of ever-improving quality pose many questions that still need to be answered. These phenomena provide clues to many aspects of 
both physics (e.g.,~interaction between plasma and radio emission, and the pulsar emission mechanism) and astrophysics (e.g.,~dynamics of 
the outflow in the binary system and the details of binary evolution).

The unusually bright pulses are one more such unexplained phenomenon.  The properties of these pulses (clustering close to eclipses, 
unchanging average pulsed intensity or profile shape during BSP outbursts, widths similar to the width of the average profile, intensities 
up to 40 times the average pulse intensity, and correlated structures in dynamic spectra spanning several pulses) make them difficult to classify 
as a separate emission mode or Giant pulses. BSPs from \T\ are similar in properties to strong pulses from B1957+10, which were attributed to strong 
lensing of the ordinary pulses by the intrabinary material \citep{Main2018}. 

In the current work we argued that strong lensing can explain, at least qualitatively, all of the aforementioned properties of \T's BSPs. 
Assuming strong lensing for \T\, and following the framework from \citet{Cordes2017} for a single 1-D lens with a Gaussian DM profile, 
results in a lens as small as 70--3400\,km  and residing as close as 0.3--800 orbital separations away from the pulsar.
The model in \citet{Main2018} yields smaller lenses closer to pulsar, with $a_\mathrm{lens}$ of 5--280\,km and $d$ of $0.06-160R_\mathrm{sep}$.
Such lenses can reside in a rarefied extended plasma tail from the companion, as shown in hydrodynamical simulations of plasma outflow in 
\citet{Tavani1991,Tavani1993}.  Those simulations predict a single tail, however, which does not explain symmetrical BSPs around eclipses, as is 
often observed for \T. In order for strong lensing to occur, the DM in the lens must be larger than $\approx 4\times10^{-5}$\,pc cm$^{-3}$. DMs obtained 
from the brightest BSPs have much larger measurement errors of about 0.1\,pc cm$^{-3}$ and are scattered around DMs from the time-averaged emission by 
the same amount. Neither DM nor the spectral index of the time-averaged emission exhibit measureable changes during BSP outbursts.

The lens size and location have been estimated based on two toy models with key parameters being known, at best, to within an order of magnitude. 
Reality is likely far more complex, and studying the properties of the BSPs provides a new and interesting way of examining the physical 
conditions in the intra-binary plasma and, potentially, giving insight into the emission region within the pulsar magnetosphere via lensing-induced 
magnification \citep{Main2018}. Also, unlike for PSR~B1957$+$10, radio emission from \T\ is considerably linearly polarized, giving a handle on the 
magnetic field in the intrabinary material via Faraday rotation and lensing properties \citep{Li2019}. 
Finally, the population of known ablating binary pulsars is growing steadily, and several may have as yet undetected BSPs.

\begin{acknowledgments} 
 AVB thanks Vlad Kondratiev for help with preparing single-pulse data, Tim Pennucci for measuring scattering time scales, Dongzi Li and Ue-Li Pen for 
 fruitful discussions.  The National Radio Astronomy Observatory is a facility of the National Science Foundation operated under cooperative agreement by Associated Universities, Inc.
\end{acknowledgments}

\newpage

\appendix
\label{app:sps_stat}

\section{A. Single-pulse and average emission properties vs. orbital phase}

\begin{figure*}
\centering 
\includegraphics[scale=1.0]{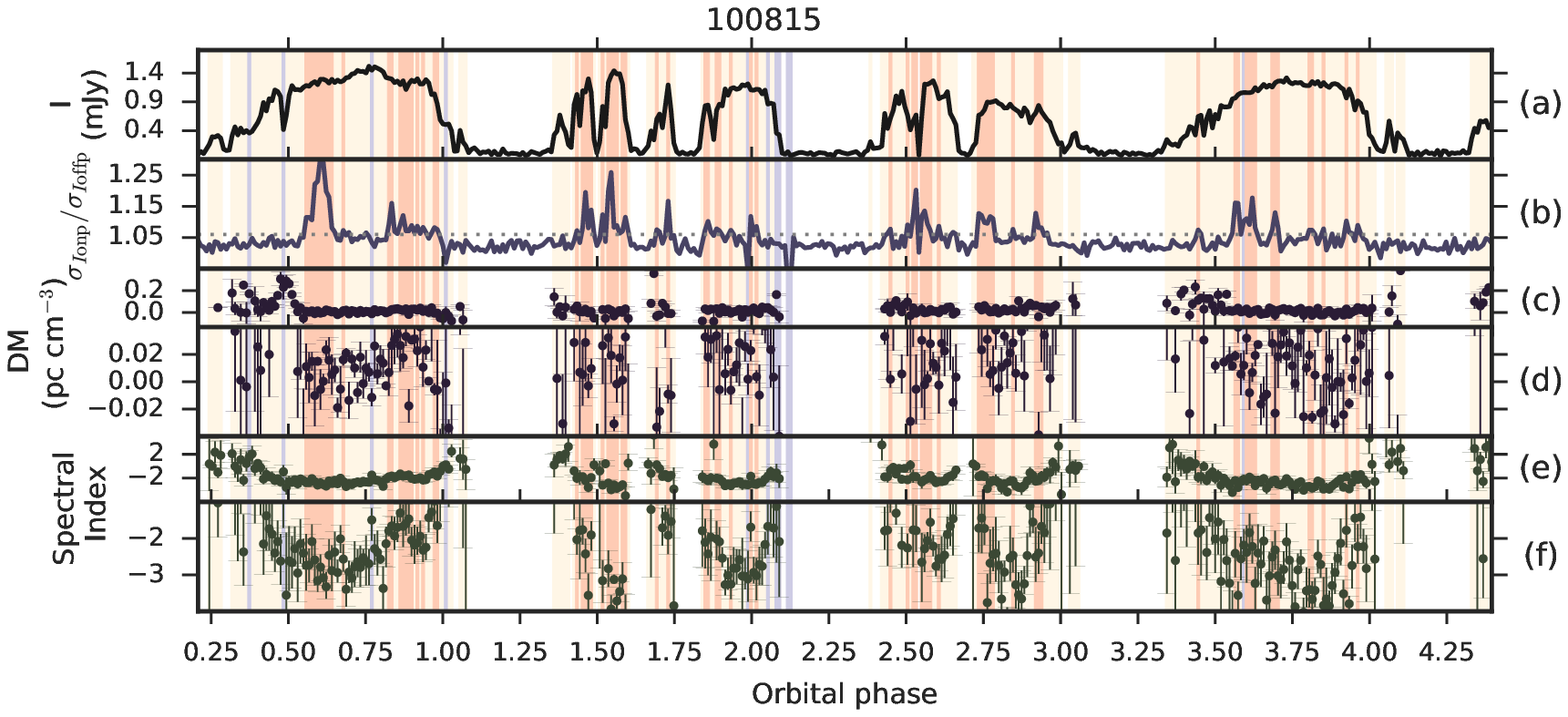}
\includegraphics[trim=-20mm 0mm 0mm 0mm, scale=0.85]{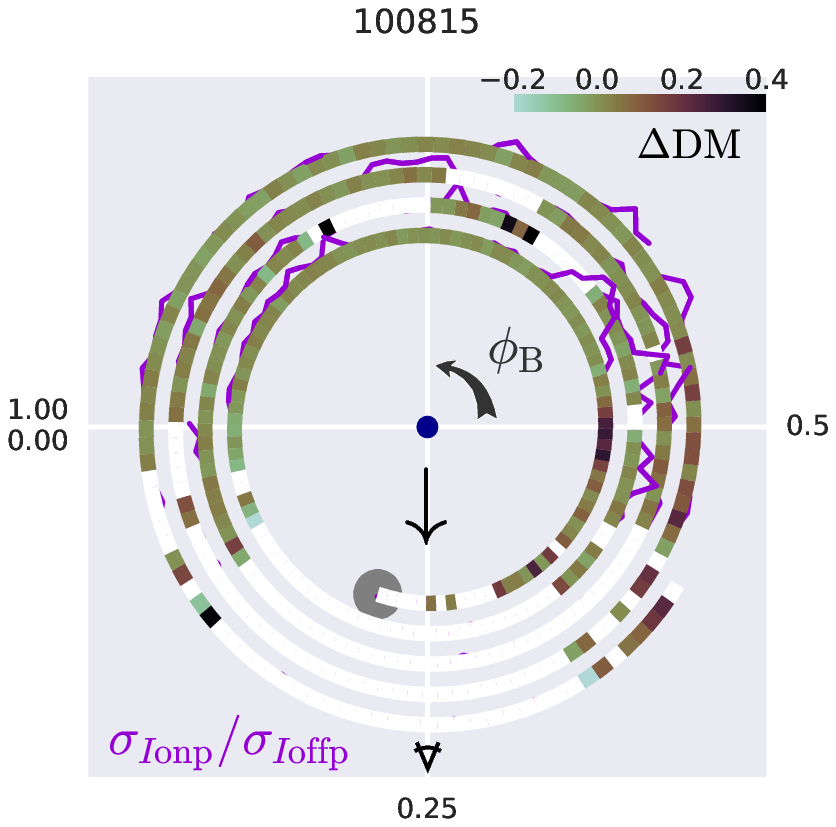}\includegraphics[scale=0.85]{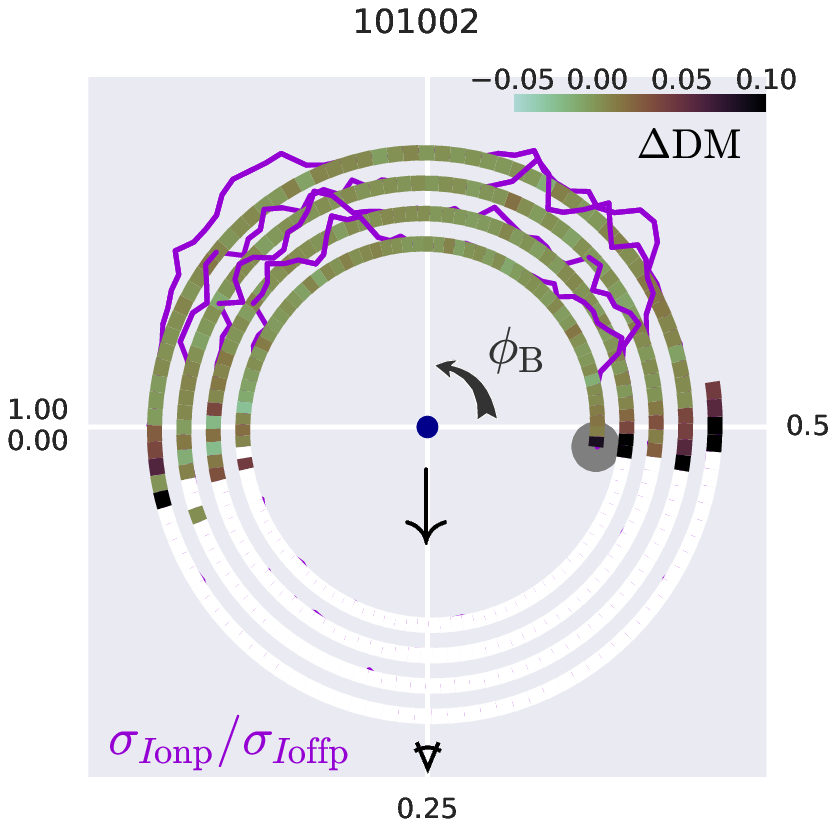}
\includegraphics[scale=1.0]{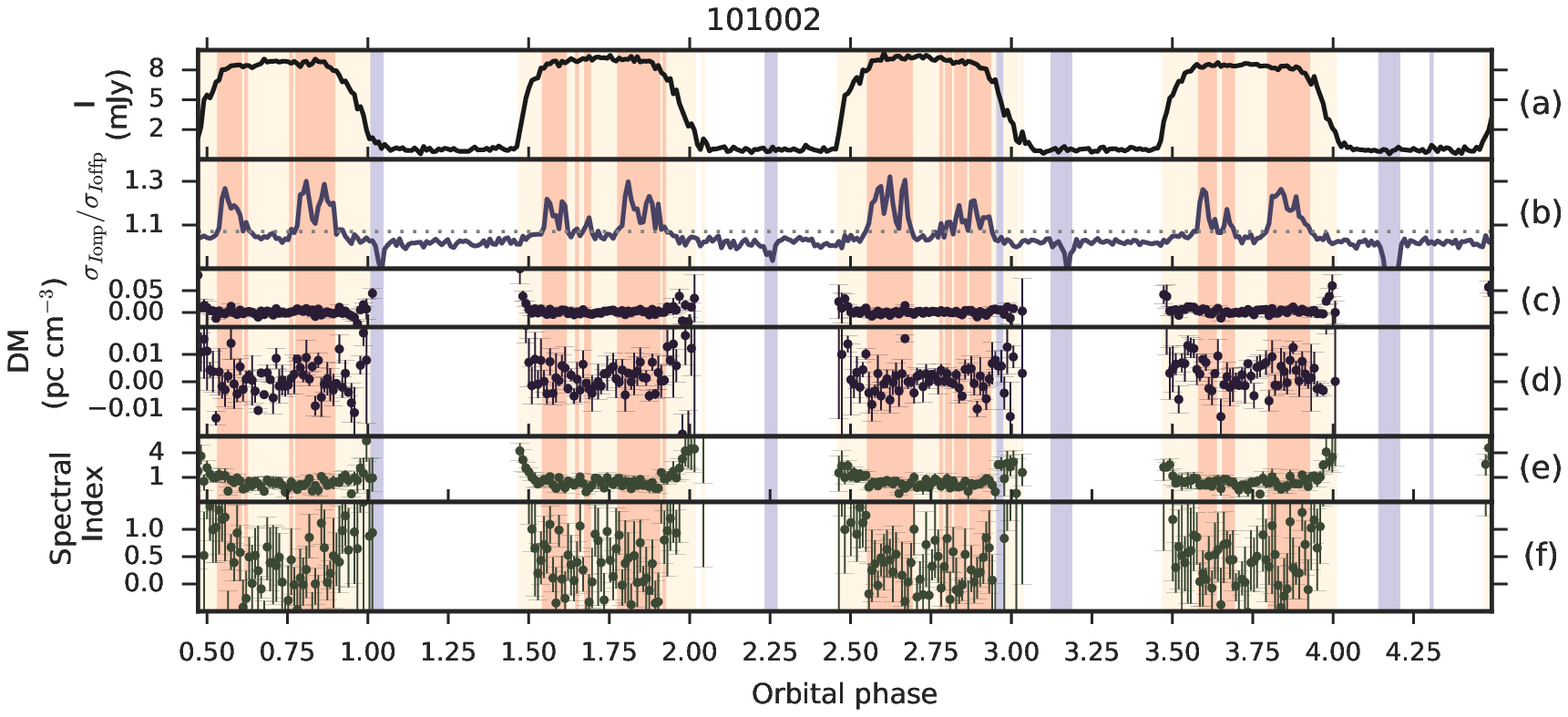}
\caption{See text for explanation }.
\label{fig:sps_stat_1} 
\end{figure*}

\begin{figure*}
\centering 
\includegraphics[scale=1.0]{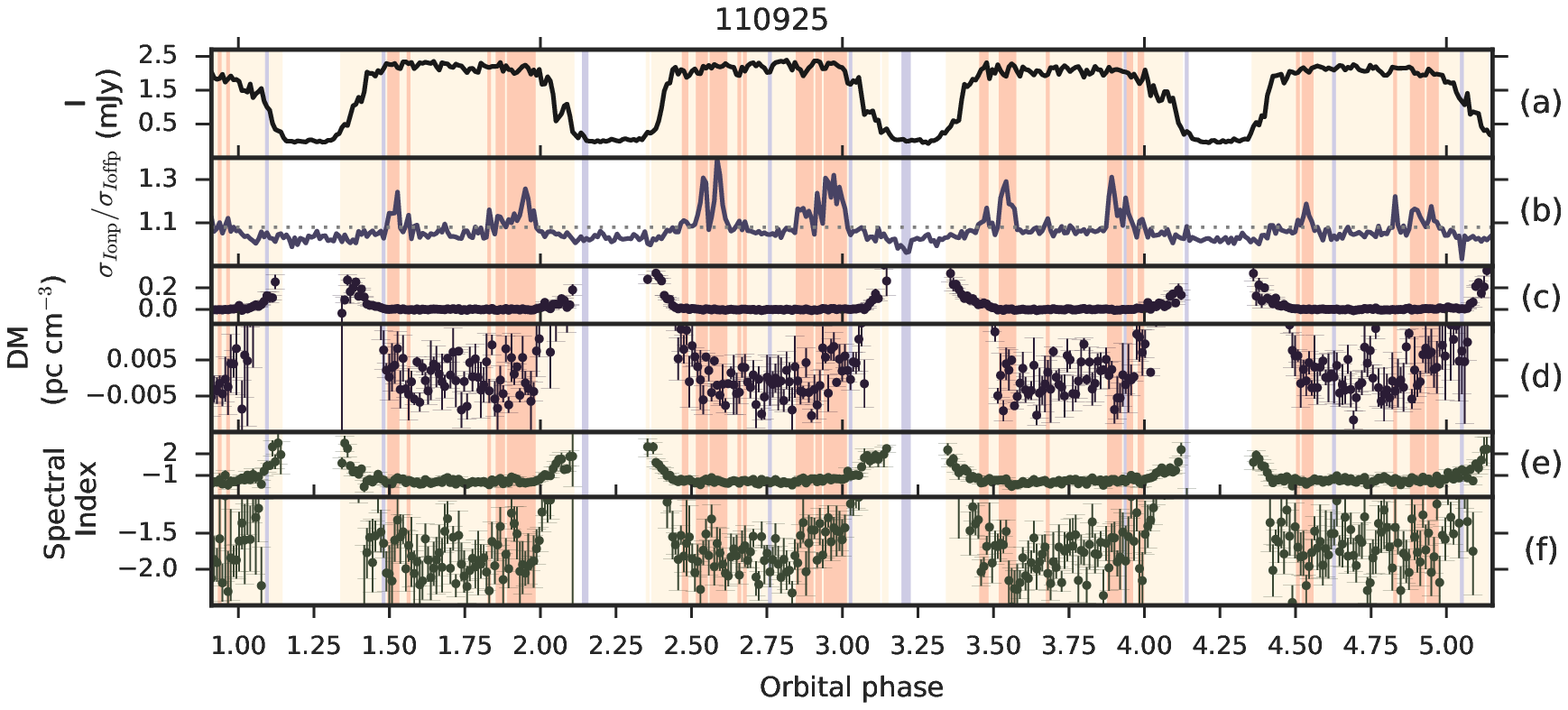}
\includegraphics[trim=-20mm 0mm 0mm 0mm, scale=0.85]{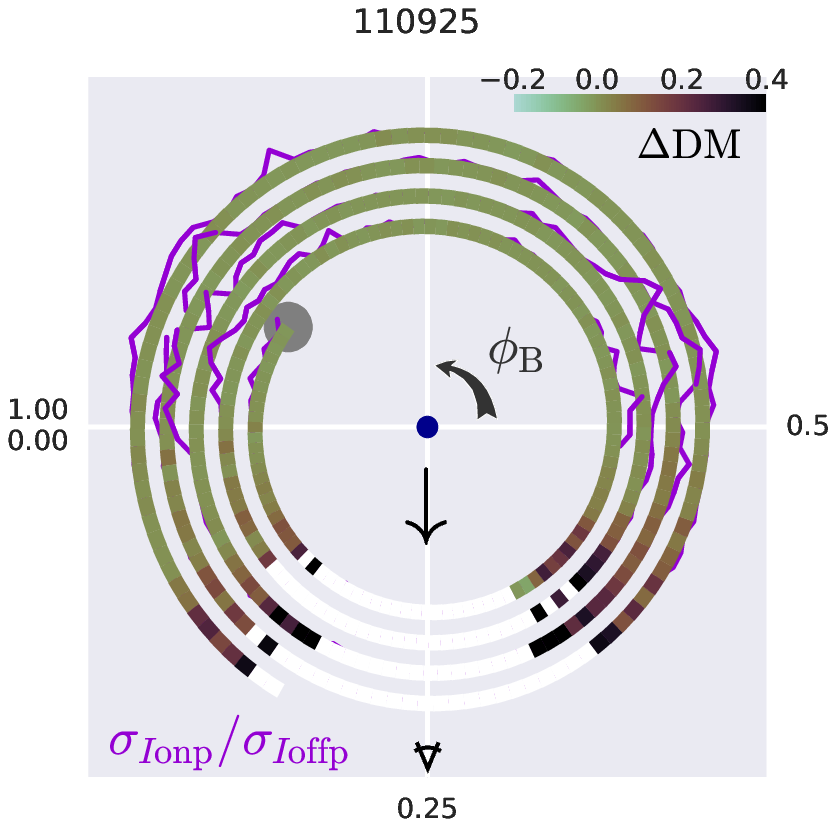}\includegraphics[scale=0.85]{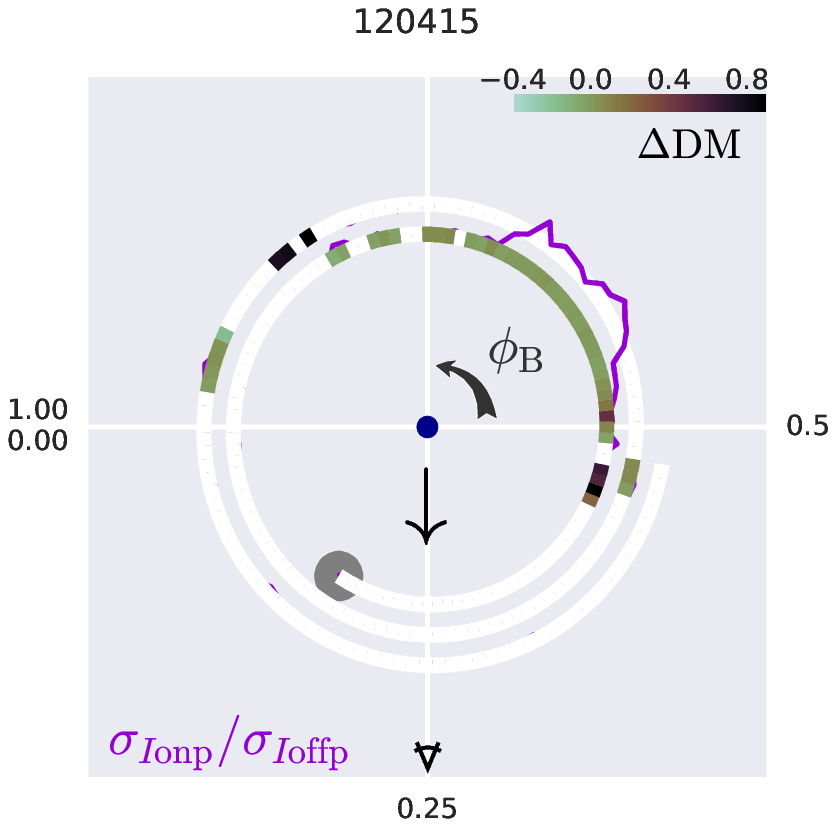}
\includegraphics[scale=1.0]{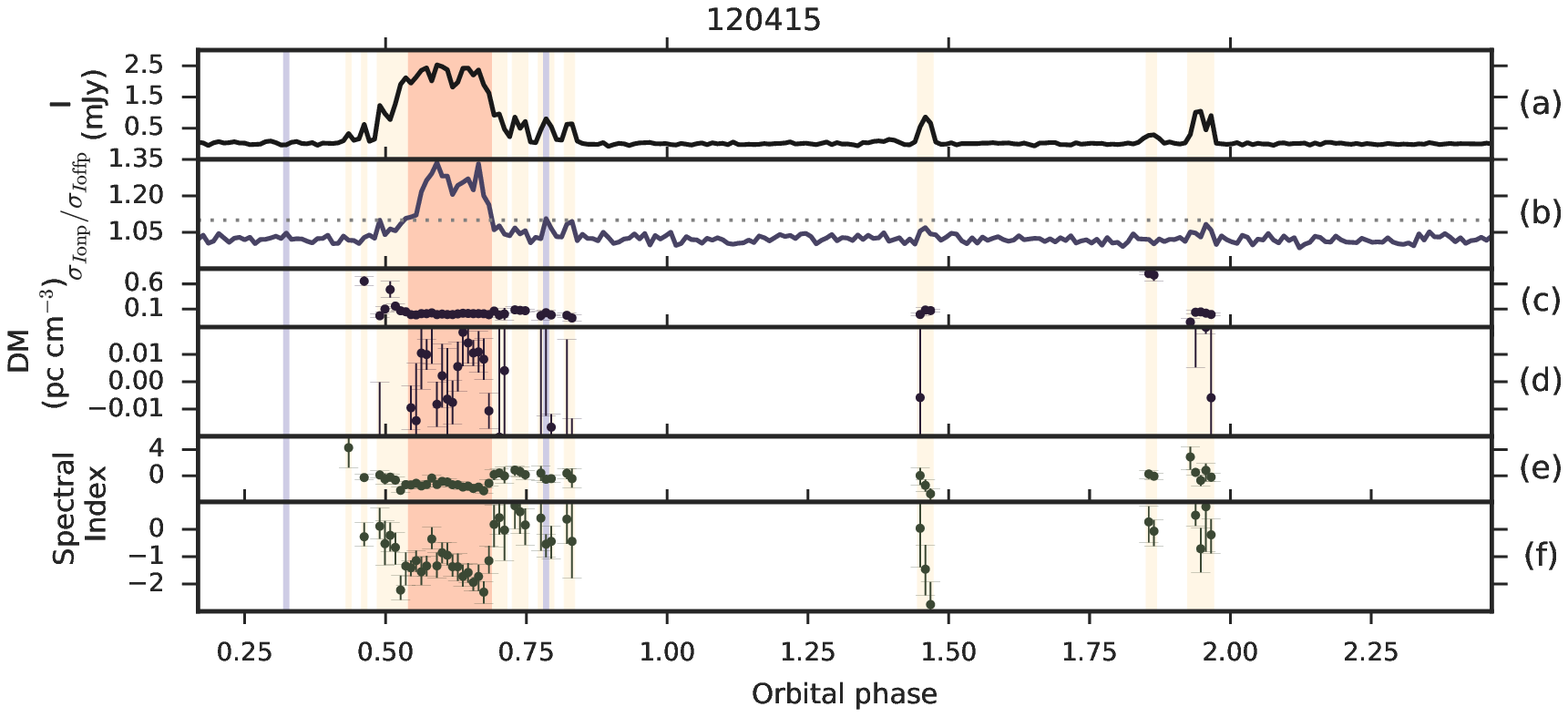}
\caption{Same as Fig.~\ref{fig:sps_stat_1}, but for sessions 110925 (top) and 120415 (bottom).}
\label{fig:sps_stat_2} 
\end{figure*}

\begin{figure*}
\centering 
\includegraphics[scale=1.0]{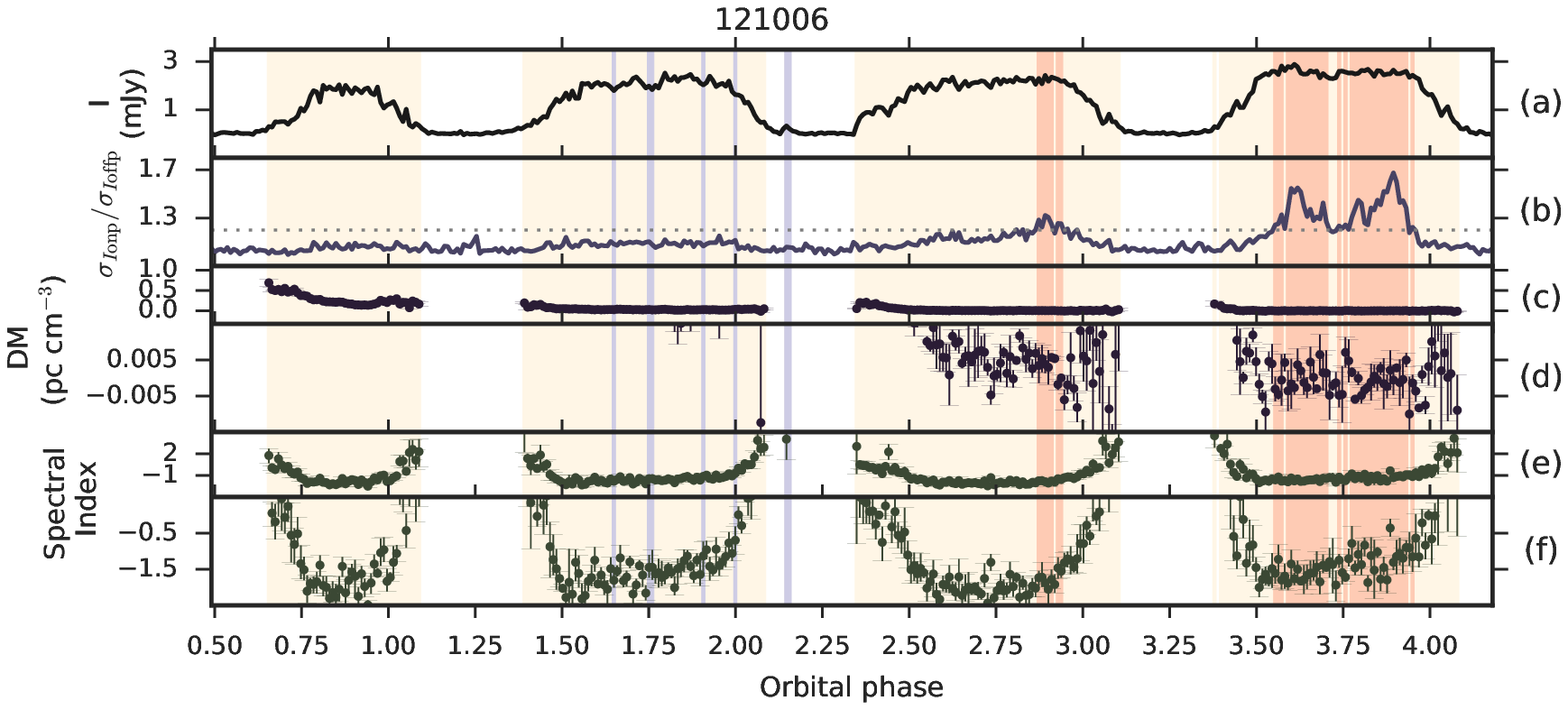}
\includegraphics[trim=-20mm 0mm 0mm 0mm, scale=0.85]{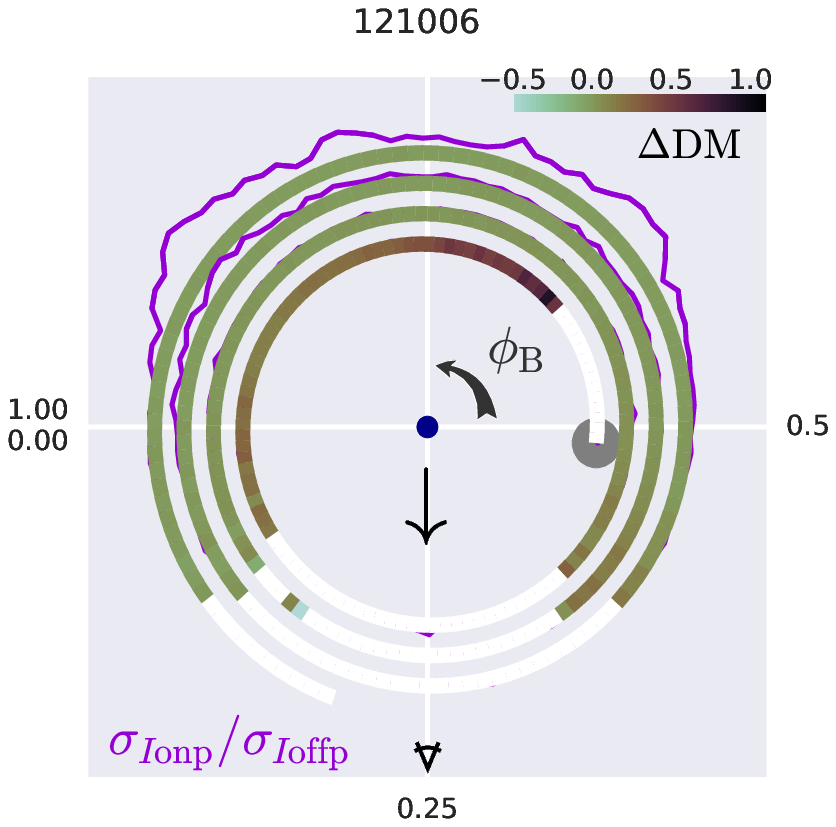}\includegraphics[scale=0.85]{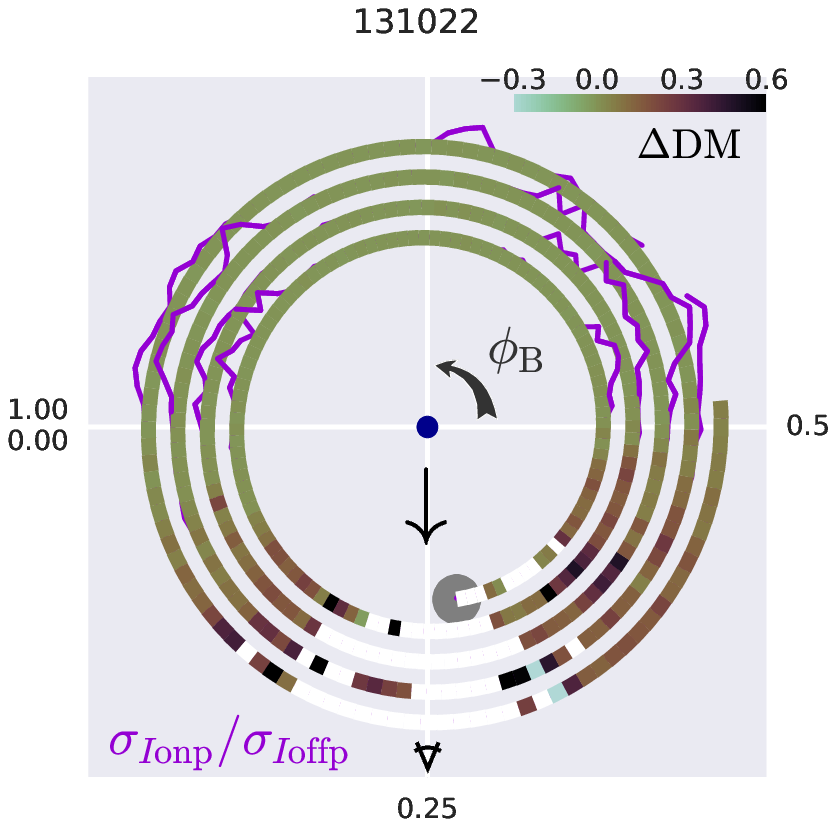}
\includegraphics[scale=1.0]{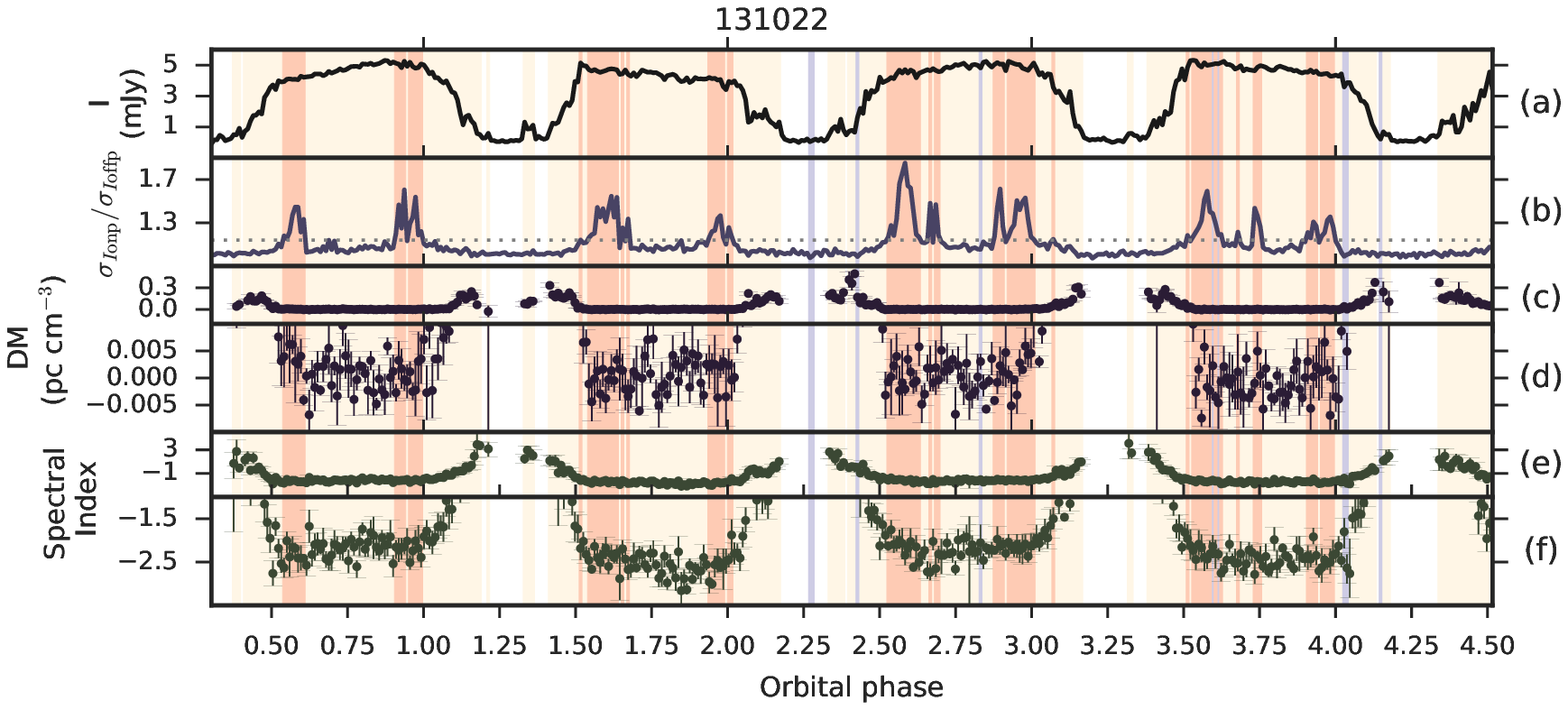}
\caption{Same as Fig.~\ref{fig:sps_stat_1}, but for sessions 121006 (top) and 131022 (bottom).}
\label{fig:sps_stat_3} 
\end{figure*}

\begin{figure*}
\includegraphics[scale=1.0]{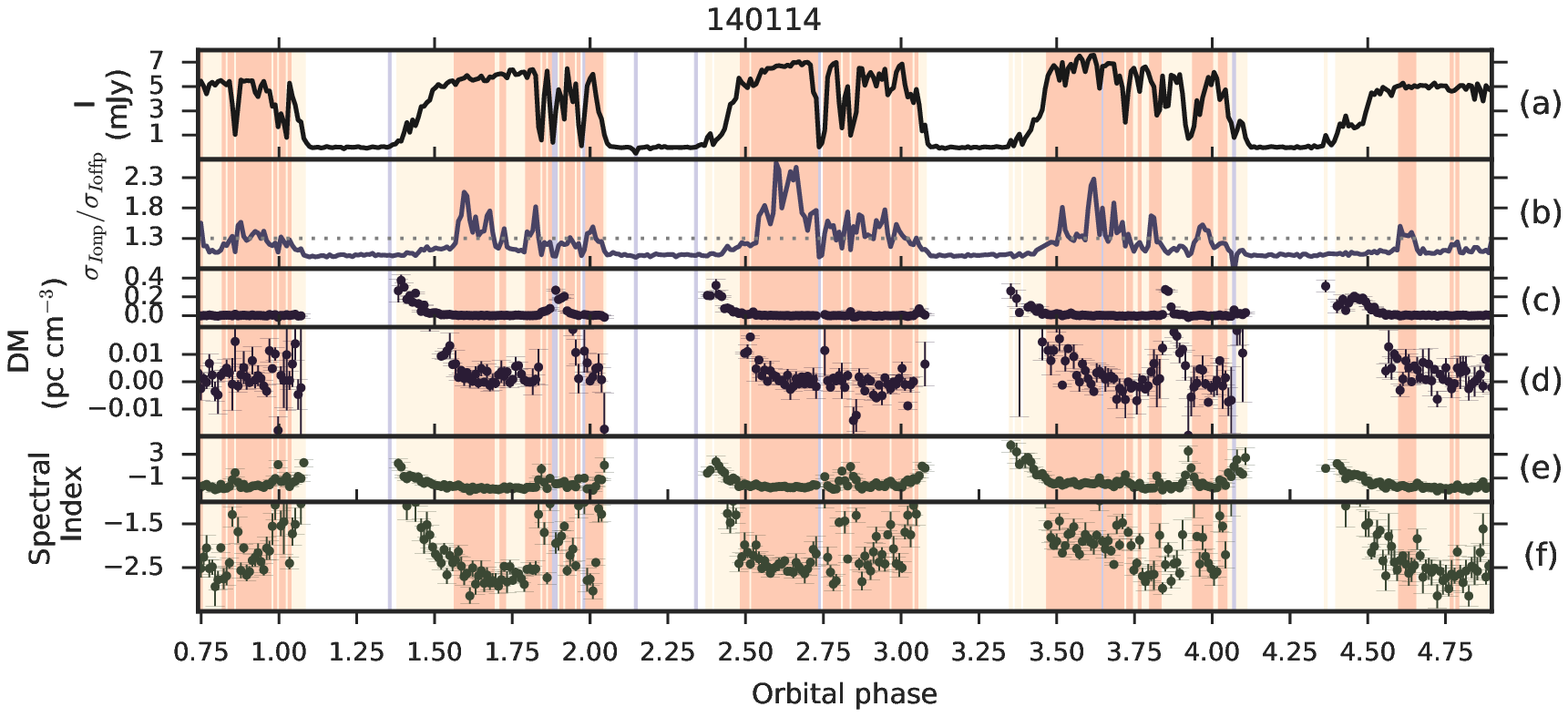}
\includegraphics[trim=-20mm 0mm 0mm 0mm, scale=0.85]{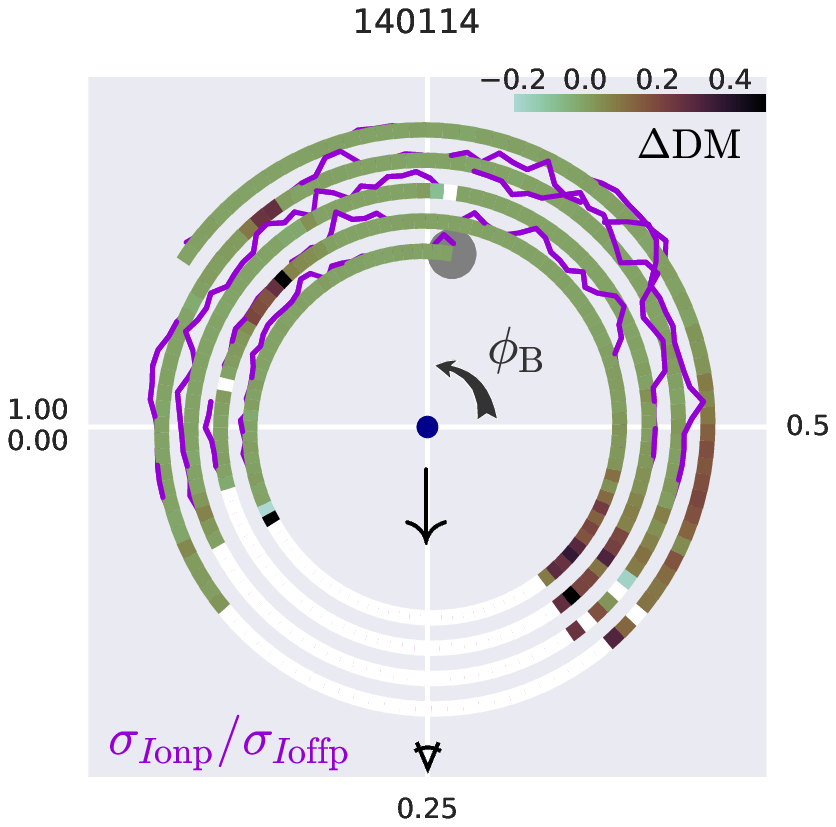}\includegraphics[scale=0.85]{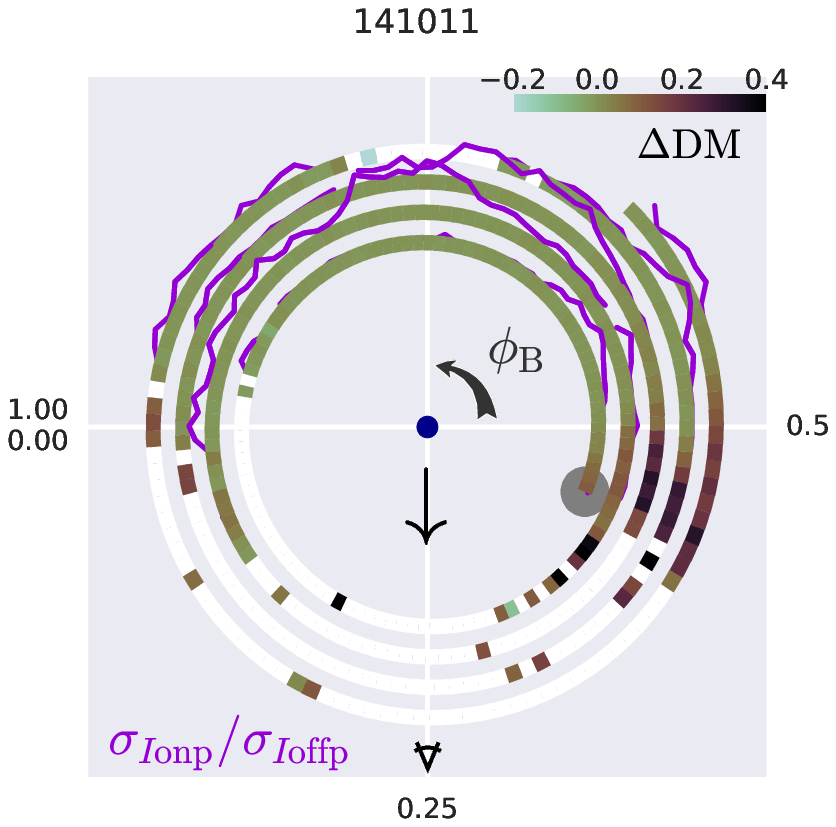}
\includegraphics[scale=1.0]{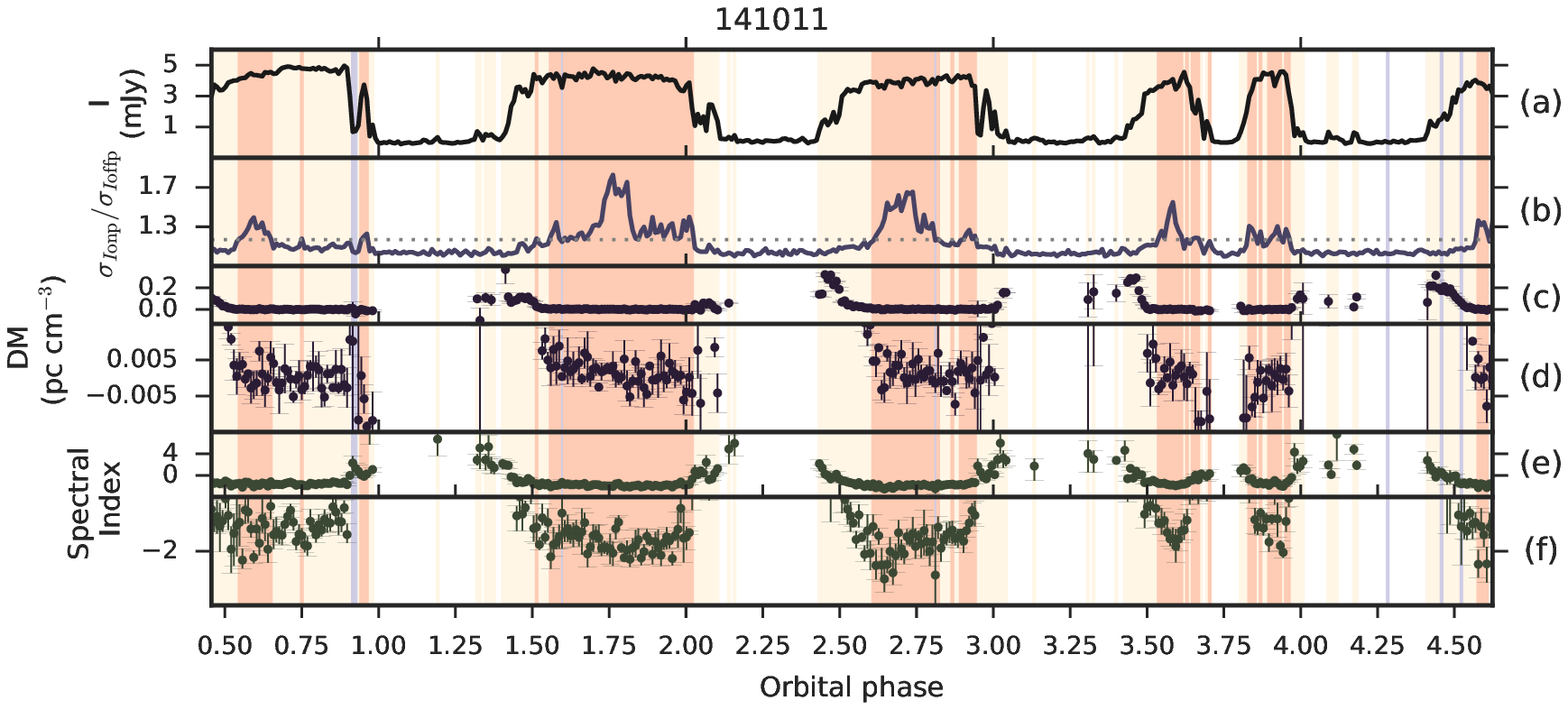}
\caption{Same as Fig.~\ref{fig:sps_stat_1}, but for sessions 140114 (top) and 141011 (bottom).}
\label{fig:sps_stat_4} 
\end{figure*}

The following section provides a description of Figs.~\ref{fig:sps_stat_1}--\ref{fig:sps_stat_4}. Each figure shows two observing sessions, 
with the observation date on top of each panel.

\subsection{Top and bottom panels}

Across all subplots: BSP, NSP, and RFI regions are shown as vertical colored stripes (brown, beige, and violet, respectively). 
In order to determine whether a particular orbital phase range belongs to one of the three aforementioned categories, we examined the standard 
deviation of the mean flux densities, $\sigma_{I_\mathrm{onp}}$, in the on-pulse phase window. The $\sigma_{I_\mathrm{onp}}$ is obviously larger 
in the $\phiB$ regions with strong pulses, however it also depends on the gradually varying SEFD and can be biased by RFI.  To eliminate the influence 
of RFI and SEFD variations, we developed the following procedure. 
First, we compute the distribution of mean flux density in the off-pulse spin phases in each of the 60-s blocks 
of data: 
\begin{equation}
I_\mathrm{offp} = \frac{1}{N_\mathrm{bin}}\sum^{\phi=0.5}_{\phi=0.38} I(\phi).   
\label{eq:Ioffp}
\end{equation}
For each 60-s block, we calculate the variance of the noise flux densities, 
$\sigma^2_{I_\mathrm{offp}} =  \langle I_\mathrm{offp}^2\rangle-\langle I_\mathrm{offp}\rangle^2$, and normalize them by subtracting the median value in a 
20-min window around each block, to compensate for SEFD variation.  Then, we calculate the mean and standard deviation of this normalized 
$\sigma^\mathrm{norm}_{I_\mathrm{offp}}$. Blocks with normalized $\sigma^\mathrm{norm}_{I_\mathrm{offp}}$ exceeding its mean by five standard deviations 
are marked as corrupted by RFI. Blocks with pulsed emission (either BSP or NSP) are defined as RFI-free if the mean on-pulse flux density value 
$\langle I_\mathrm{onp}\rangle$ exceeds $\langle I_\mathrm{offp}\rangle$ by $5\sigma_{I_\mathrm{offp}}$. Finally, the BSP and NSP blocks are separated 
by  $\sigma_{I_\mathrm{onp}}/\sigma_{I_\mathrm{offp}}$ equal to the manually set, session-dependent threshold ranging from 1.06 to 1.3. 
For each session, the exact value of threshold was chosen by examining the variation of $\sigma_{I_\mathrm{onp}}/\sigma_{I_\mathrm{offp}}$ 
during orbital phase regions with high enough average flux densities and without visible DM increase.

Individual subplots:

\noindent(a): The average flux density in the on-pulse phase region per pulse block (calculated according to Eq.~\ref{eq:Ionp}).

\noindent(b): The ratio of the standard deviation of the flux density values in the onpulse/offpulse regions, with the dotted horizontal line marking the BSP threshold.

\noindent(c): DM measured with TOAs produced by cross-correlating the average profile in 60-s integrations and four subbands with a single template. 
The template was obtained by averaging the data from the highest-S/N session in the given band, excluding orbital phases affected by excess dispersion. 
If S/N of the signal was low in some of the subbands (mostly happening in the vicinity of eclipses, where frequency-dependent attenuation is large), 
then the corresponding TOA was deleted. In the vicinity of the eclipses, DM values may be biased by profile changes due to scattering.

\noindent(d): Same as (c), but zoomed to the typical DM values in the orbital phase regions further from eclipses.

\noindent(e): Spectral index of the 60-s folded data, with 8 subbands per band. The spectra were well fit by a power-law. Flux density values were 
calculated in fixed on-pulse windows, and therefore when close to eclipses, spectral indices may be biased by excess dispersion and scattering. Interestingly, 
the spectral index is flat or positive in all orbital phases in the UHF band, indicating a spectral turnover somewhere around 820\,MHz. Such a high frequency 
for the spectral turnover is rare (usually, pulsars exhibit turnover around 100\,MHz or lower), but not unique:  \T\ may be another example of so-called pulsars 
with GHz-peaked spectra \citep{Kijak2011}.

\noindent(f): Same as (e), but zoomed to the typical spectral index values in the orbital phase regions further from eclipses.

\subsection{Middle panel}

Spatial distribution of DM and BSP regions with respect to the pulsar/companion positions and the LOS. The center of mass of the system is shown in the 
center of the plot, with the pulsar orbit (projected semimajor axis of 0.12\,lt-s) being located within the blue marker. The grey circle shows the Roche 
lobe of the companion (0.35\,lt-s) at the separation of 2.5\,lt-s \citep{Nice1992}. Orbital phase increases counterclockwise. The LOS is marked with an 
arrow and observer with an eye pictogram. Both the pulsar and companion are circling around the center of mass, with trajectory of the companion plotted 
as an unwinding spiral to show multiple orbits. Orbital phase of zero is set at the ascending node. Although it is ruled out by the occasional presence 
of pulsar radiation throughout eclipses, the orbit is drawn edge-on. 

For each orbital phase, the color of the spiral marks the DM along the LOS. Thus, DM at orbital phases around 0.75 does not represent the electron 
density behind the pulsar, but DM along LOS when the companion is behind the pulsar. The range of color corresponds to the range on subplots (c) 
of the panels above or below, with darker color indicating larger DM.  White indicates the absence of pulsed emission 
or the insufficient S/N for a reliable DM measurement.

The dark violet line plotted along the spiral shows $\sigma_{I_\mathrm{onp}}/\sigma_{I_\mathrm{offp}}$ from the subplots (b) of the upper or 
lower panels. The line corresponds to the standard deviation of the single pulses traveling along the LOS. Interestingly, BSP regions tend 
to emerge when the companion is at orbital phase of about 0.6 or 0.8 (but not necessarily, see, for example, sessions 140114 and 141011). 

\newpage

\section{B. NSP and BSP flux density distributions}
\label{app:efit}

\begin{figure*}
\centering 
\includegraphics[scale=0.92]{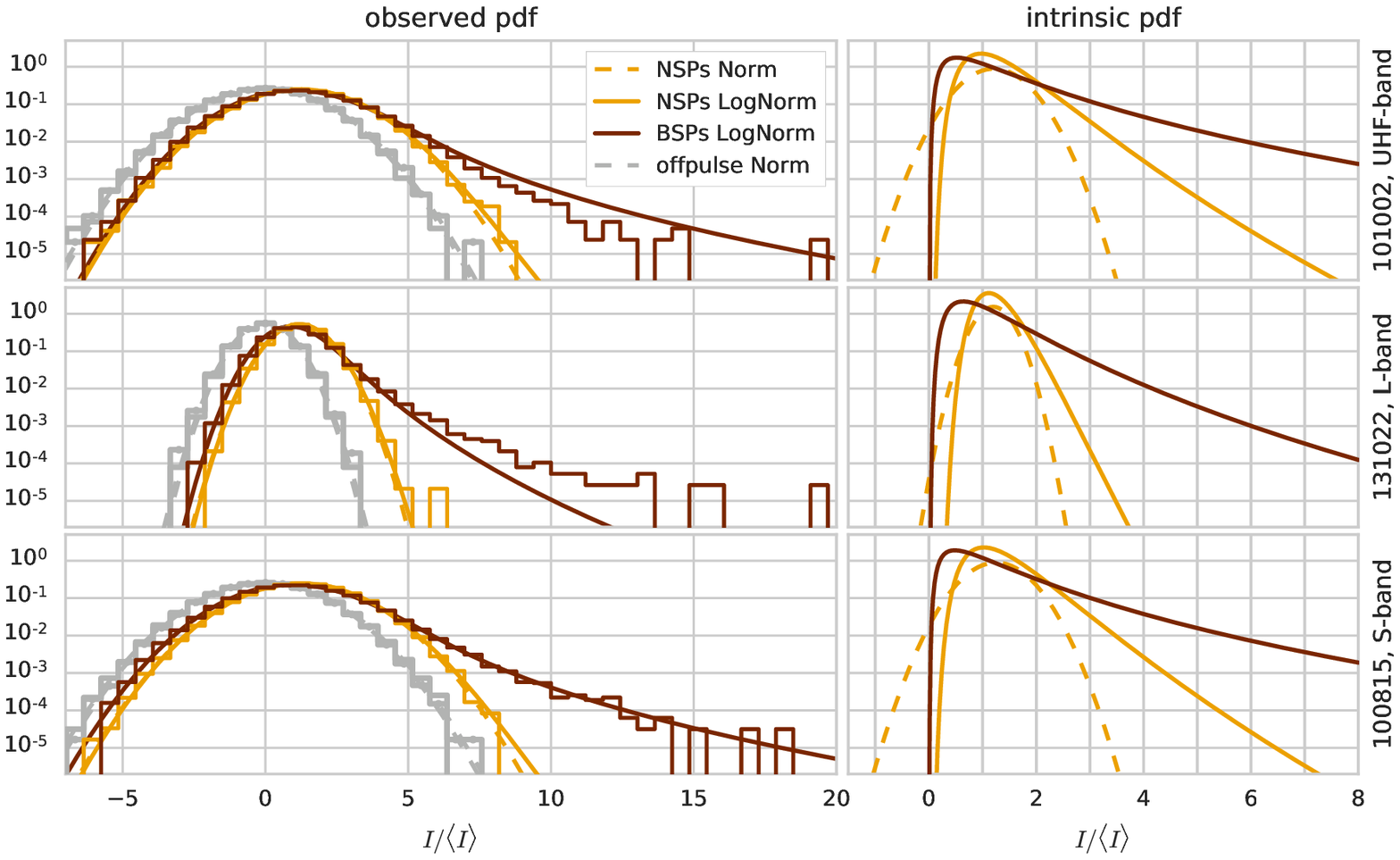}
\caption{\textit{Left column:} single-pulse flux density distributions (normalized by the average uneclipsed flux density, Table~\ref{table:obssum}) for 
the NSP (orange histogram) and BSP (brown histogram) orbital phase regions for the three sessions representing three observing bands (Fig.~\ref{fig:AP}). 
For the session 131022, the strongest pulse had $I\approx30\langle I\rangle$ (not shown). The gray histograms 
(two for each session, for the BSP and NSP orbital phase regions)
mark the flux density distributions of noise and the dashed gray lines show corresponding normal (i.e.,~gaussian) fits. Orange and brown solid 
lines show the convolution of a best-fit intrinsic log-normal distribution with the noise fit, whereas the dashed orange lines show the same 
convolution, but with the best-fit intrinsic normal distribution. The best-fit intrinsic distributions are shown in the \textit{right column}.}
\label{fig:Edistr} 
\end{figure*}

\begin{table*}
\begin{center} 
\caption{Flux density distribution fits\label{table:efit}}
\def\arraystretch{2}\tabcolsep=2pt
\begin{tabular}{l|cc|c|c|c|} 
\cline{2-6}
& \multicolumn{3}{c}{$\frac{}{}$NSP} & \multicolumn{2}{|c|}{BSP} \\ 
\cline{2-6}
 & \multicolumn{2}{c|}{$P_\mathrm{intr}$} & $P_\mathrm{offp}$ & $P_\mathrm{intr}$ & $P_\mathrm{offp}$ \\
\cline{1-6}
\multicolumn{1}{|c|}{} & LogNorm & Norm & Norm & LogNorm & Norm \\
\multicolumn{1}{|c|}{Session}& $(\mu_\mathrm{LN}, \sigma_\mathrm{LN})$ & $(\mu_\mathrm{N}, \sigma_\mathrm{N})$ & $(\mu_\mathrm{N}, \sigma_\mathrm{N})$ & $(\mu_\mathrm{LN}, \sigma_\mathrm{LN}$ & $(\mu_\mathrm{N}, \sigma_\mathrm{N})$ \\ 
\hline
\multicolumn{1}{|l|}{101002, UHF} & (0.05, 0.17) & (1.21, 0.45) & (0.00, 1.52) & (-0.04, 0.33) &  (0.01, 1.49) \\
\multicolumn{1}{|l|}{131022, L-band}  & (0.07, 0.10) & (1.20, 0.26) & (-0.00, 0.71) & (-0.05, 0.25) & (-0.00, 0.69) \\
\multicolumn{1}{|l|}{100815, S-band}  & (0.07, 0.16) & (1.25, 0.45) & (-0.01, 1.54) & (-0.07, 0.33) & (-0.01, 1.57) \\
\hline 
\end{tabular} 
\end{center}
\end{table*}

The flux density distribution of NSPs and BSPs in each observing band was estimated based on sample $\sim10$\,min-long trains of pulses from
sessions 101002, 131022, and 100815 (Fig.~\ref{fig:AP}). The flux densities in on- and off-pulse regions (calculated according to Eq.~\ref{eq:Ionp} 
and Eq.~\ref{eq:Ioffp}, respectively) were normalized by $\langle I\rangle$, the average flux in the corresponding session (Table~\ref{table:obssum}). 
The probability density function of $I_\mathrm{onp}/\langle I\rangle$, $P_\mathrm{onp}$ is the convolution between the intrinsic pulse flux density 
distribution, $P_\mathrm{intr}$, and the distribution of $I_\mathrm{offp}/\langle I\rangle$, $P_\mathrm{offp}$:
\begin{equation}
 P_\mathrm{onp} = P_\mathrm{intr}\otimes P_\mathrm{offp}
\end{equation}
In order to find $P_\mathrm{intr}$, we first approximated $P_\mathrm{offp}$ 
as the probability density function of the normal distribution:
\begin{equation}
\mathrm{Norm}(\mu_\mathrm{N}, \sigma_\mathrm{N}) \sim \frac{1}{\sigma_\mathrm{N}\sqrt{2\pi}}\exp\left[-\frac{(I/\langle I\rangle-\mu_\mathrm{N})^2}{2\sigma_\mathrm{N}^2}\right].
\end{equation}

The offpulse flux density distribution appears to be well fit with the normal distribution (Fig.~\ref{fig:Edistr}), although no formal evaluation of the goodness of fit was performed.

The intrinsic flux density distribution was modeled either with normal (NSPs) or log-normal (NSPs, BSPs) distributions. Following \citet{Burke-Spolaor2012}, 
we chose the following parameterization of a log-normal distribution:
\begin{equation}
\mathrm{Lognorm}(\mu_\mathrm{LN}, \sigma_\mathrm{LN}) \sim \frac{1}{I/\langle I\rangle\sigma_\mathrm{LN}\sqrt{2\pi} }\exp\left[ -\frac{[\log_{10}(I/\langle I\rangle)-\mu_\mathrm{LN}]^2}{2\sigma_\mathrm{LN}^2} \right].
\end{equation}
The convolution of $P_\mathrm{intr}$ and the best-fit $P_\mathrm{offp}$ was fitted to the observed $P_\mathrm{onp}$ by minimizing the sum of squared 
difference between the model and data in flux density bins with at least one pulse. The choice of model functions is to a large extent arbitrary. 
Clearly, the lognormal function does not approximate the flux density distribution of BSPs well, failing to reproduce the high-energy tail of one of the sessions. 
More elaborate models (perhaps including the influence of propagation effects) would result in better fits.

Best-fit values of the mean and standard deviation for all three bands and two choices of model distribution are listed in Table~\ref{table:efit}. 
Apparent negative intrinsic flux densities for the normal fits in UHF and S-bands stem from the intrinsic flux density being small and the fit being 
contaminated by noise. For normal fits, $\mu_\mathrm{N}$ is larger than one because of the fortuitous choice of the $\phiB$ window where the local 
mean flux density was larger than the mean flux density per observation, $\langle I\rangle$.

\bibliographystyle{apj} 
\bibliography{Ter5A_bibliography}

\end{document}